\documentclass{emulateapj}

\newcommand{\Htwo}{H$_2$}
\newcommand{\NHI}{$N_{\rm HI}$}
\newcommand{\HST}{{\it HST}}
\newcommand{\FUSE}{{\it FUSE}}
\newcommand{\kms}{km~s$^{-1}$}
\newcommand\etal{et~al.}
\newcommand{\cd}{cm$^{-2}$}

\begin{document}

\title{Metallicity and Ionization in High Velocity Cloud Complex C}    

\author{Joseph A. Collins\altaffilmark{1}, J. Michael Shull}

\affil{University of Colorado, CASA, Department of Astrophysical 
\& Planetary Sciences, Campus Box 389, Boulder, CO 80309 \\ 
jcollins@casa.colorado.edu, mshull@casa.colorado.edu}

\altaffiltext{1}{Also at Front Range Community College, Larimer Campus, 
4616 S. Shields St., Fort Collins, CO 80526}

\and

\author{Mark L. Giroux}
\affil{East Tennessee State University, Department of Physics, Astronomy,
   \& Geology \\ Box 70652, Johnson City, TN 37614 \\ girouxm@etsu.edu}

\begin{abstract}

We analyze \HST\ and \FUSE\ ultraviolet spectroscopic data for eleven
sight lines passing through the infalling high velocity cloud (HVC) 
Complex~C. These sight lines pass through regions with H~I column
densities ranging from \NHI\ = $10^{18.1}$ to $10^{20.1}$ \cd. 
From [\ion{O}{1}/\ion{H}{1}] abundances, we find that Complex~C 
metallicities range from 0.09 to 0.29 $Z_{\sun}$, with a column density 
weighted mean of $0.13~Z_{\sun}$. Nitrogen (\ion{N}{1}) is underabundant 
by factors of (0.01--0.07)(N/H)$_{\sun}$, significantly less than oxygen 
relative to solar abundances.  This pattern suggests nucleosynthetic 
enrichment by Type~II SNe, consistent with an origin in the Galactic 
fountain or infalling gas produced in winds from Local Group galaxies.  
The range of metallicity and its possible ($2\sigma$) dependence on \NHI\ 
could indicate some mixing of primordial material 
with enriched gas from the Milky Way, but the mixing mechanism is unclear.  
We also investigate the significant highly ionized component of Complex C, 
detected in \ion{C}{4}, \ion{Si}{4}, and \ion{O}{6}, but not in \ion{N}{5}.  
High-ion column density ratios show little variance and are consistent 
with shock ionization or ionization at interfaces between Complex C 
and a hotter surrounding medium.  Evidence for the former mechanism is 
seen in the Mrk 876 line profiles, where the offset in line centroids 
between low and high ions suggests a decelerating bowshock.

\end{abstract}
\keywords{Galaxy: halo --- ISM: clouds --- ISM: abundances --- 
quasars: absorption lines}

\section{Introduction}

The origin of the objects known as high-velocity clouds (HVCs; Wakker
\& van Woerden 1997) remains a problem within the framework of Galaxy
evolution.  Traditionally studied through \ion{H}{1} 21-cm emission,
HVCs do not have a detectable stellar component, and they are characterized
by line centroids, typically at negative local standard of rest velocities,
$V_{LSR}$, that deviate from
models of Galactic rotation.  The subsolar metallicities of HVCs imply
that some fraction of the gas consists of primordial material.  The
continuous infall of low-metallicity ($\sim0.1 Z_{\sun}$) gas onto the
disk is predicted by Galaxy formation models that seek to explain such
phenomena as the G-dwarf problem (Pagel 1994).  Although their
distances are uncertain, HVCs may trace structure formation on Galactic 
or Local Group scales.

One of the best studied HVCs in recent years has been Complex~C,
infalling gas located at least 6 kpc distant (Wakker 2001).
Consideration of its large angular size ($\sim2000$ deg$^{2}$)
effectively limits it to a nearby location in the Galactic halo.
Complex C is pierced by more UV-bright quasar sight lines than any
other high-velocity cloud, making it an ideal laboratory to
investigate the nature of high-velocity gas.  Ion abundances have been
investigated in several studies (Wakker \etal\ 1999, Gibson \etal\ 2001,
Richter \etal\ 2001, Collins \etal\ 2003 [hereafter CSG03], Tripp \etal\ 
2003), all of which indicate that the complex has a subsolar
metallicity.  The most robust of these studies utilize the
[\ion{O}{1}/\ion{H}{1}] abundance\footnotemark
\footnotetext{Throughout this work, for element $X$, this quantity is
defined as, [$X$/H I]=log$\left(\frac{N(X)/N(HI)_{\rm
C}}{A(X)_{\sun}}\right)$, where $A(X)_{\sun}$ is the solar abundance
of element X.  The solar abundances for several elements have changed
considerably in the last few years (AGS05).} to determine metallicity.  
Since \ion{O}{1} is coupled
to \ion{H}{1} through charge exchange, we can assume that
[\ion{O}{1}/\ion{H}{1}] $\approx$ (O/H).  Other ion abundances such as
[\ion{S}{2}/\ion{H}{1}] require substantial and uncertain ionization
corrections to determine the elemental abundance.  By investigating
all Complex C {\it Far Ultraviolet Spectroscopic Explorer} (\FUSE)
and {\it Hubble Space Telescope} (\HST) 
sight lines available at the time, CSG03 derived a metallicity range, 
$Z=0.1-0.25~Z_{\sun}$, based on the three sight lines for which 
[\ion{O}{1}/\ion{H}{1}] could be measured.  

This range of metallicities suggests that Complex~C is a mixture of 
infalling primordial material (with $Z \approx 0.1~Z_{\odot}$) and 
higher-metallicity material that is produced in the Galactic disk and 
then elevated in a ``Galactic fountain" (Shapiro \& Field 1976; Bregman 
1980).  Alternatively, such metal-poor gas might be expelled from
Local Group dwarf galaxies into the Galactic halo. 
We hope to gain insight into how these two reservoirs entrain and mix 
by observing elemental abundances and ionization distributions in  
multiple sight lines through Complex~C. 
Further evidence for an enrichment of Complex C by Galactic material
comes from analysis of elemental abundances, whose
pattern can yield information about the nucleosynthetic
history of the constituent metals (Pettini 2004).  Alpha-process
elements (such as O, Si, and S) are produced predominantly by Type II
supernovae (SNe), while Fe is produced primarily by Type Ia SNe.
Nitrogen is produced mostly by lower mass stars in the
asymptotic giant branch (AGB) stage.  With the bulk of
$\alpha$-element production occurring relatively soon after
star-formation is initiated, the release of N and Fe into the ISM can
lag S or O enrichment by $\sim250$ Myr (Henry \etal\ 2000).  Many 
of the previous Complex C studies find that N is depleted
significantly relative to the $\alpha$-elements, sometimes by as much
as a factor of 10.  Several studies also found evidence for a
slight decrement in Fe that cannot be attributed to dust depletion.  
CSG03 suggest that this abundance pattern is consistent with enrichment 
by Type II SNe, followed by ejection from the star-forming region in a 
Galactic fountain before significant N and Fe enrichment.  

If the infalling Complex C is mixing with fountain material, there should 
be observable signatures of that mixing.  If mixing occurs more
efficiently in the outer regions of the complex, while cloud cores
remain relatively pristine, then one may expect to see an
anti-correlation between component metallicity and $N_{\rm HI}$.  Based on the
three sight lines with accurate [\ion{O}{1}/\ion{H}{1}] metallicities,
CSG03 found evidence for such a dependence, with the highest column
density sight line at the lowest metallicity.  However, the statistics of 
the correlation are poor, and measurements of additional sight lines 
are needed.  

Since the low ions track elemental abundances, Complex C is typically
discussed in those terms.  However, the cloud has multiphase structure
consisting of highly ionized gas, detected in \ion{C}{4}, \ion{Si}{4}, 
and \ion{O}{6} (Fox \etal\ 2004; hereafter F04).  In previous work 
(Collins, Shull, \& Giroux 2004, 2005, hereafter CSG04 and CSG05).
discussed the presence of
multiphase gas, produced in bow shocks produced at the interfaces
between infalling HVCs in the Galactic halo.  The H$\alpha$ emission 
measures of Complex C and other HVCs are consistent
with photoionization by escaping radiation from OB associations in the
disk (Bland-Hawthorn \& Putman 2001).  CSG03 found that the low-ion
abundance pattern in Complex C is consistent with such an ionization
source.  Although photoionization can explain the production of 
neutral and singly-ionized species, it cannot simultaneously
explain the highly ionized component.  F04 analyzed high ion ratios
for the PG~1259+593 sight line and found that they are consistent with
the production of highly ionized gas at the interface of Complex C and
a hotter surrounding medium.  Since Complex C is a known halo object
with abundant quasar sight lines, it provides a test case for studying
the production of highly ionized gas in the general HVC population
(Sembach \etal\ 2003).  The understanding of the production of highly
ionized high-velocity gas is of particular interest for the study of
the unusual class of ``highly ionized HVCs'' (Sembach \etal\ 1999;
Collins, Shull, \& Giroux 2004, 2005, hereafter CSG04 and CSG05). 
We have suggested (CSG04, CSG05) that the majority of highly ionized HVCs 
represent low column density analogs to objects such as Complex C,
whereas Nicastro \etal\ (2002, 2003) have proposed that they trace 
shock-heated filaments of the warm-hot intergalactic medium (WHIM) at 
large distances.  Better characterization of the highly ionized component 
of Complex C may shed light on the nature of the highly ionized HVCs.

Since the publication of CSG03, several new Complex~C data sets from \FUSE\ 
and \HST\ appeared. These new datasets fall into one of three categories: 
1) new or additional \FUSE\ data that have
increased the effective exposure time, in some cases significantly; 
2) new \HST-STIS E140M, E230M, or G140M exposures; 
3) Mrk 205, a Complex C sight line of low-\NHI\ with high quality \FUSE\ 
and \HST-STIS E140M data neglected in
previous Complex C studies.  Among these new datasets are \FUSE\ 
spectra for 7 Complex C sight lines, including \HST-STIS data for 5 of 
these 7 sight lines.   We couple these new data with previously presented
Complex C data for the 8 sight lines from CSG03 and one (3C~351) from 
Tripp \etal\ (2003).  We update all previous column densities and
abundances, using recent solar abundances (AGS05) and revised 
absorption oscillator strengths (Morton 2003).  
With this sample, we investigate issues of Complex C 
metallicity, gas mixing, relative abundances, and high ion production.  
The \FUSE\ and \HST\ observations are reductions discussed in \S\ 2.  
The results for the analysis for the new sight lines are presented in 
\S\ 3.  Finally, in \S\ 4 we discuss and interpret the results.

\section{Description of Observations and Reductions}

\begin{deluxetable*}{lccccc}
\tablecolumns{8}
\tablewidth{0pc}
\tablecaption{Summary of Newly Analyzed Observations\tablenotemark{a} \label{t1}}
\tablehead{
\colhead{} & \colhead{FUSE} & \colhead{FUSE} &  \colhead{HST-STIS} & \colhead{HST} & \colhead{HST-STIS}\\
\colhead{Sightline} & \colhead{Program ID} & \colhead{$T_{\rm exp}$(ks)\tablenotemark{b}} & 
\colhead{Grating} & \colhead{Program ID} & \colhead{$T_{\rm exp}$(ks)}}
\startdata
HS 1543+5921  & P108             & 8.5   & G140M  & 9784 & 25.3, 54.0, 49.5\tablenotemark{c} \\
Mrk 205       & Q106, S601, D054 & 203.6 & E140M  & 8625 & 78.3 \\
Mrk 279       & P108, D154       & 181.8 & E140M  & 9688 & 41.4 \\
Mrk 290       & D076, E084       & 68.1  & E230M  & 8150 & 32.5 \\
Mrk 501       & P107, C081       & 31.7  & ...    & ...  & ...  \\
Mrk 876       & P107, D028       & 129.1 & E140M  & 9754 & 29.2 \\
PG 1626+554   & P107, C037       & 98.5  & ...    & ...  & ...  
\enddata
\tablenotetext{a}{Including only sight lines that either have data yet to be 
published for a Complex C study or have additional data since publication of CSG03.}
\tablenotetext{b}{Effective exposure time for the LiF1a channel.}
\tablenotetext{c}{G140M observations are taken at three separate grating tilts. 
The three listed exposure times are for central wavelengths of 1222~\AA, 1272~\AA, 
and 1321~\AA, respectively.}
\end{deluxetable*}

The new Complex C sight line data for this investigation include
\FUSE\ and \HST-STIS datasets that were either previously unanalyzed or  
taken since the
work for CSG03 was completed.  All datasets were obtained from the
publicly available Multimission Archive (MAST) at the Space Telescope
Science Institute.  Of the eight sight lines from CSG03, five have new
\FUSE\ or \HST-STIS datasets, including Mrk 279, Mrk 290, Mrk 501, Mrk
876, and PG 1626+554.  Also included here are \FUSE\ and \HST-STIS data
for the sight lines HS 1543+5921 and Mrk 205, previously unanalyzed 
in the context of Complex C.  Table 1 summarizes the observations
that have new or unanalyzed \FUSE\ and \HST-STIS data.

Seven Complex C sight lines have new \FUSE\ data.  For a complete 
description of the \FUSE\ instrument and its operation see Moos \etal\ 
(2000) and Sahnow \etal\ (2000).  All sight lines were observed through 
the $30\arcsec
\times30\arcsec$ LWRS aperture in time-tag mode, except for the new
C037 dataset for PG 1626+554, which was observed through the $4\arcsec
\times20\arcsec$ MDRS aperture.  Owing to the small aperture size of the
MDRS data, the target is not present in the PG 1626+554 SiC data.
Calibrated \FUSE\ data were extracted using a pass through the CALFUSE
version 2.4 reduction pipeline.  In order to improve signal-to-noise
for these data, we include both ``day'' and ``night'' photons in the
final calibrated spectra.  The inclusion of ``day'' photons leads to
strong airglow contamination of interstellar \ion{O}{1}\ and
\ion{N}{1}\ absorption lines.  However, the contamination is generally
centered at $V_{LSR}=0$ \kms\ and does not affect the
higher-velocity Complex C absorption.  Individual exposures were
coadded, weighted by their signal-to-noise, to yield a final spectrum.
Before coadding datasets taken several years apart, we checked spectra 
for unusual differences between datasets.  If the differences
were judged to be severe, then offending datasets were not included in
the final coadded spectrum.  In several cases, significant deviations
were discovered between datasets taken at different times.  The most
significant difference was a factor of $\gtrsim5$ decrease in the flux
of Mrk 279 in its 2002 data.  The 2002 data are not included in our final
Mrk 279 spectrum.  Other less significant differences warranting
rejected data for specific LiF or SiC segments include the absence of
segment data or a strong continuum deviation.

A single resolution element of the \FUSE\ spectrum is about 
20 \kms\ ($\sim10$ pixels) and, as a result 
data are oversampled at that resolution.  Thus, to further improve 
signal-to-noise, we rebinned the data over 5 pixels.  
To set the absolute wavelength scale, we compared the centroid of 
Galactic \ion{H}{1}\ 21-cm emission and aligned it to various Galactic 
absorption lines in the \FUSE\ 
bandpass, such as those of \ion{Si}{2}\ ($\lambda$1020.70), 
\ion{O}{1}\ ($\lambda$1039.23), \ion{Ar}{1}\ ($\lambda\lambda$1048.22, 
1066.66), \ion{Fe}{2}\ ($\lambda\lambda$1125.45, 1144.94), \ion{N}{1}\
($\lambda$1134.17), and various H$_{2}$ Lyman bands. We estimate the 
absolute wavelength scale to be accurate to within $\sim10$ \kms.

The new \HST-STIS data consist of observations taken with the E140M, E230M, or 
G140M gratings.  The E140M and E230M echelle modes 
provide for a resolution of 7 \kms\ over 1150--1700 \AA\ and 10 \kms\ 
over 1850--2700 \AA, respectively.  The 
HS 1543+5921 dataset is taken with the first-order G140M grating, covering 
the wavelength 
range 1194--1348 \AA\ at a resolution of $\sim25$ \kms.  Final
spectra were obtained by coadding individual exposures, weighted by
their signal to noise.  In order to both improve signal-to-noise and
to match pixel size in velocity to that of the rebinned \FUSE\ 
data, we re-binned the E140M and E230M echelle data to 3 pixels, while
no rebinning was used for the first-order G140M spectrum.  Finally, we
set the absolute wavelength scale as for the \FUSE\ data,
comparing the \ion{N}{1}\ ($\lambda$1199.550) and \ion{S}{2}\
($\lambda$1250.584, 1253.811, 1259.519) lines to the centroid of
Galactic \ion{H}{1}\ emission.

\section{Measurements From New Data}

In this section, we present measurements of ion column densities from
the new data for Complex C sight lines.  In each of these sight lines,
except Mrk 501, we detect metal-line absorption associated with
Complex C.  The Mrk 501 data exhibit features that are likely due to
a low-\NHI\ Complex C component, although the low signal to noise in
that data prevents measurements greater that $3\sigma$ significance
for any of the Complex C features.  As a result, the Mrk 501 sight line
provides only weak upper limits on abundances, and we do not discuss
the sight line further.

In order to measure column densities of ion species, we begin by
extracting individual line profiles from the full \FUSE\ and \HST-STIS
spectra.  These profiles are normalized by fitting low-order
polynomials to the continuum $\pm$3--10~\AA\ about the line in question,
although in a number of cases spurious features near the line required
the use of a much larger region for continuum measurement.  For each
sight line, we determine the velocity range of Complex C from the
\ion{H}{1}\ profile.  We then measure the equivalent width, $W_{\lambda}$, 
of an absorption feature by integrating the line over the
velocity range occupied by Complex C.  This technique works well for
low ions (neutral or singly-ionized), but high-ions (\ion{C}{4},
\ion{Si}{4}, \ion{O}{6}) can extend in velocity by as much as 45 \kms\ 
beyond the bounds established by the \ion{H}{1} profile.  In those cases, 
we integrate the high ions over the full observed high-velocity
absorption.  When lines cannot be measured, we estimate 3$\sigma$
upper limits to $W_{\lambda}$.  

The \FUSE\ bandpass contains numerous \Htwo\ lines that often
contaminate high-velocity absorption for metal lines.  In cases of
weak contamination of high-velocity features, we can often quantify
the contribution from contaminating \Htwo.  In order to model the
contribution from a contaminating \Htwo\ line, we analyze other
isolated \Htwo\ lines of the same rotation state $J$.  By fitting 
several lines for a given $J$ with Gaussian profiles, and fitting measured
values of $W_{\lambda}$ with a curve of growth, we can model the
expected profile of the contaminant line.  If the contamination of the
Complex C feature is weak, we can subtract the contaminating absorption 
to reveal the Complex C profile and measure its equivalent
width.  An example of such a fit is shown in Figure 1 to the
\ion{O}{1} $\lambda1039.230$ line for Mrk 290.  As indicated by the
\Htwo-subtracted profile, the remaining \ion{O}{1} feature resembles other
Complex C absorption lines seen in that sight line 
(see Figure 5 for other Mrk 290 profiles).  In
many cases, the \Htwo\ contamination is too strong to accurately analyze
the underlying Complex C profile (note the \Htwo\ contamination of high-velocity
\ion{O}{1} $\lambda1039.230$ in the Component 2 profile for Mrk 876 
in Figure 6).

\begin{figure}
\figurenum{1}
\epsscale{1}
\plotone{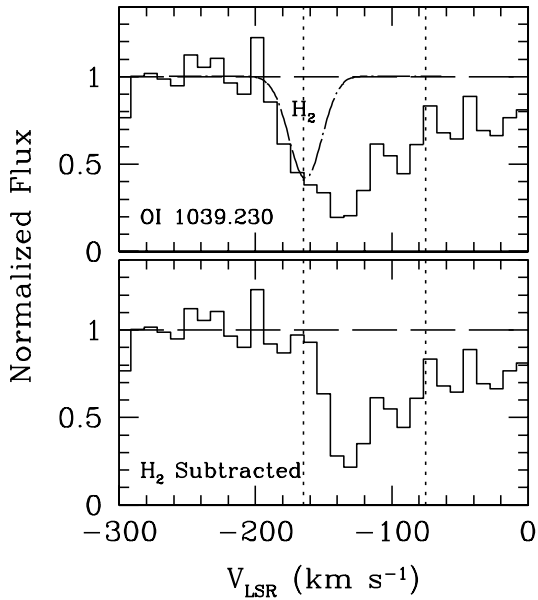}
\caption{An example of the \Htwo\ fitting procedure to contaminated
Complex C lines.  Shown is the normalized profile of high-velocity
absorption for the \ion{O}{1} $\lambda1039.230$ line in the data of 
Mrk~290, along with the modeled profile of the Galactic L5-0 R(2) \Htwo\ 
line at 1038.69~\AA\ ({\it top}).  Also shown is the profile with the
\Htwo\ model subtracted ({\it bottom}).  The clean subtraction and the
profiles similarity to those of other ions (as in Figure 5) indicates
a successful subtraction.  The vertical dashed lines indicate the
$-165$ to $-75$ \kms\ range of integration for Complex~C as 
determined by the \ion{H}{1} profile.}
\end{figure} 

In order to determine the column densities of ion species, we attempt
to fit the data with a curve of growth.  This requires the measurement
of minimally two lines for one species in any given Complex C
component.  We are able to empirically fit a curve of growth to low
ions in Complex C components towards Mrk 205, Mrk 279, Mrk 290, and
Mrk 876.  In cases where a curve of growth cannot be fit, we use the
apparent optical depth (AOD) method to measure column densities of
lines that are clearly optically thin (Savage \& Sembach 1991).  In
other cases, where unresolved saturation may be an issue, as 
occurs for the \ion{O}{1}~$\lambda1039.23$ line in particular, we 
estimate a range of possible
column densities.  Since the primary goal of this work is to
specifically determine \ion{O}{1} abundances, we estimate the column
density range by using the observed spread in doppler $b$-values
determined from curve-of-growth fits to data including 
\ion{O}{1} lines in other Complex C sight lines.
Considering all of the Complex C sight lines that we have analyzed
here and in CSG03, the $b$ values fitted to data that include \ion{O}{1} 
lines range from 10.2 \kms\ $\leq b \leq 18.0$ \kms.  Using this spread of 
$b$-values, we can establish the range of possible column densities 
from lines of moderate optical depth.  The high-ions \ion{C}{4}, \ion{Si}{4},
\ion{N}{5}, and \ion{O}{6} probably reside in a different component than 
the low ions, and are thus not well described by a $b$-value determined
from species of a lower degree of ionization.  For the high-ions,
we use the apparent optical depth method to measure column
densities.  All rest wavelengths and oscillator strengths are from
Morton (2003).

Values of \NHI\ are measured from data taken at the Effelsberg 100-m
telescope, except in the case of HS 1543+5921, for which we measure \NHI\ 
from data of the Leiden-Dwingeloo Survey (LDS; Hartmann \& Burton 1997).  
We adopt the Effelsberg \NHI\ values quoted in Wakker \etal\ (2003), 
except in the cases of Mrk 279 and Mrk 876, which exhibit blended
multi-component Complex C structure.  In those cases, we directly
integrate the Effelsberg profile to calculate \NHI.  The beam sizes of
the Effelsberg telescope and LDS are $9\arcmin.1$ and $36\arcmin$, 
respectively, whereas the beam size in
quasar absorption spectroscopy is effectively determined by the
sub-arcsecond angular size of the quasar emitting region.  Therefore,
the measured values sample column densities on a much larger
scale than the metal ion column densities.  Wakker \etal\ (2001)
compare \NHI\ for targets measured through both Ly$\alpha$ absorption
in \HST\ data with \ion{H}{1} emission in Effelsberg data.  They find
that \NHI\ varies by as much as 25\% between these two techniques.  We
therefore adopt a systematic error for \NHI\ measurements of 0.1 dex.
This uncertainty becomes relevant when we determine ion abundances in
\S\ 4.

\subsection{HS 1543+5921}

HS 1543+5921 is an extragalactic target that was observed with the \HST-STIS 
G140M grating for a program investigating an intervening low-surface-brightness
galaxy (Bowen et al. 2005).  The target has a low UV flux 
($\sim2\times10^{-15}$ ergs s$^{-1}$ cm$^{-2}$ \AA$^{-1}$) 
and long exposures 
were taken at three different grating tilts covering the wavelength range 
1194$-$1348 \AA.  A short 8 ks \FUSE\ exposure exists for this sight line, 
but due to the low flux of the target, those data are of insufficient quality 
for absorption line studies.

\begin{deluxetable}{lrlcc}
\tablecolumns{5}
\tablewidth{0pc}
\tablecaption{SUMMARY OF MEASUREMENTS--HS 1543+5921 SIGHTLINE\label{t2}}
\tablehead{
\colhead{} & \colhead{$\lambda$\tablenotemark{a}} & \colhead{} & \colhead{$W_{\lambda}$\tablenotemark{b}} & \colhead{log $N(X)$\tablenotemark{b}}  \\
\colhead{Species} & \colhead{(\AA)} & \colhead{$f$\tablenotemark{a}} & \colhead{(m\AA)} & \colhead{($N$ in cm$^{-2}$)} }
\startdata
\ion{H}{1}  & ...      \ \ & \ \ ...                     & ...        & $19.67^{+0.05}_{-0.07}(\pm 0.1)$\tablenotemark{c} \\
\ion{N}{1}  & 1199.550 \ \ & \ \ 0.132                   & $151\pm24$ & $[14.23^{+0.15}_{-0.14},14.92^{+0.49}_{-0.42}]$                \\
\ion{N}{5}  & 1238.821 \ \ & \ \ 0.156                   & $<47$      & $<13.33$                \\
\ion{S}{2}  & 1259.518 \ \  & \ \ 0.0166                 & $42\pm9$   & $14.30^{+0.08}_{-0.10}$                \\
...         & 1253.805 \ \  & \ \ 0.0109                 & $27\pm9$   & ...                \\
\enddata
\tablenotetext{a}{Wavelengths and oscillator strengths from Morton (2003).}
\tablenotetext{b}{Equivalent widths for Complex C integrated over velocity 
range $-170$ to $-105$ \kms.  Error bars on measured column densities are 
$1 \sigma$; upper limits are 3$\sigma$. Range of column densities 
for \ion{N}{1} is calculated from curves of growth with doppler 
parameters, $b=10.2$ \kms\ and $b=18.0$ \kms\ (in brackets). 
All other lines use the AOD method, with \ion{S}{2} column density 
based on 1259.518 \AA\ line only.} 
\tablenotetext{c}{The \NHI\ column density includes a systematic error 
(in parentheses) of 0.1 dex because of the beam-size mismatch between 
\ion{H}{1} 21-cm data and quasar absorption spectra.}
\end{deluxetable}

\begin{figure*}
\figurenum{2}
\epsscale{1}
\plotone{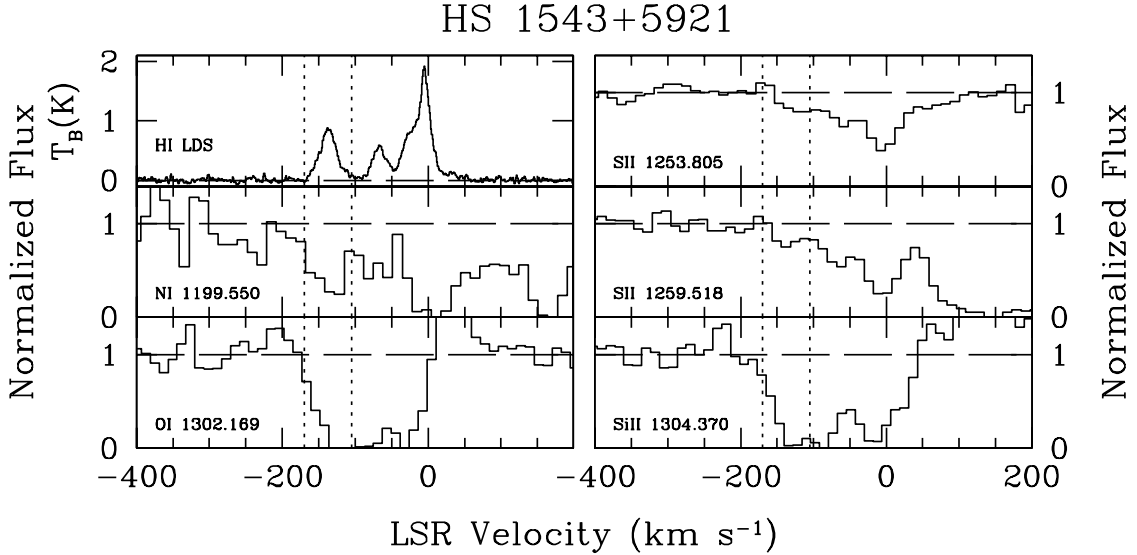}
\caption{A sample of normalized absorption profiles from \HST\ 
STIS G140M data for HS~1543+5921, along with the LDS profile of 
\ion{H}{1} emission.  The vertical dashed lines indicate the $-170$ to 
$-105$ \kms\ range of integration for measurements of $W_{\lambda}$.}
\end{figure*} 

Several of the prominent absorption lines observed in the available
STIS bandpass are shown in Figure 2 along with the LDS profile of
\ion{H}{1} emission.  The \ion{H}{1} profile shows a prominent Complex
C component that we measure at
log~$N$(\ion{H}{1}) $=19.66^{+0.03}_{-0.05}$ from direct integration of
the LDS profile.  Based on the observed line profiles, we adopt a
Complex C integration range of $-170<V_{LSR}<-105$ \kms.  The
Complex C component is easily detected in \ion{N}{1}~$\lambda1199.55$,
\ion{Si}{3}~$\lambda1206.50$, \ion{S}{2}~$\lambda\lambda1253.81, 1259.52$,
\ion{O}{1}~$\lambda1302.17$, and \ion{Si}{2}$~\lambda1260.42$.  Owing to
saturation and blending with a strong intermediate-velocity component,
we are unable to properly measure column densities of \ion{O}{1} and
\ion{Si}{2} based on the profiles for O~I~$\lambda 1302$ and 
Si~II~$\lambda1304$ presented in Figure 2.  Only the
\ion{S}{2} and \ion{N}{1} lines are sufficiently distinguishable from
lower velocity absorption for equivalent width and column density
measurements.  Equivalent widths of measured species are shown in
Table 2.

Although we have measurements for two \ion{S}{2} lines, they provide no 
constraint on the doppler parameter since they are both clearly optically thin 
lines.  Since there are insufficient absorption lines for a good curve-of-growth
analysis, we utilize the AOD method for determining column densities
in cases where the profiles are clearly optically thin.  AOD
measurements of the \ion{S}{2} lines give values of log~$N$(\ion{S}{2})
that are within 0.01 dex of one another.  We adopt the measurement
from the 1259.52 \AA\ line, based on its cleaner and deeper profile.
The \ion{N}{1}~$\lambda1199.55$ Complex C feature is quite strong, and
unresolved saturation may be an issue.  For that line, we establish a
range of possible \ion{N}{1} column densities based on the range of
observed \ion{O}{1}-based $b$ values (10.2 \kms\ $<b<$ 18.0 \kms) from
our full Complex C sight line sample.  Measured column densities are
reported in Table 2.  Owing to the strength of the 
\ion{N}{1}~$1199.55$ line in Complex~C, 
the range of possible \ion{N}{1} column densities
extends over a factor of 5.  This low end of the range represents the
nearly optically thin case, while the upper end of the range
represents the highest level of saturation seen in similar strength
lines in our Complex C sample.

\subsection{Mrk 205}

The Mrk 205 sight line passes through a thin, yet detectable,
\ion{H}{1} column of Complex C.  From Effelsberg \ion{H}{1} data,
Wakker et al. (2003) report a value of
$N$(\ion{H}{1})$=(1.3\pm0.3)\times10^{18}$ cm$^{-2}$ for the Complex C
component in this sight line, the lowest \NHI\ of the Complex C regions in
this study.  Owing in part to its low column density, this sight line
has yet to be investigated thoroughly in previous Complex C studies.

There are currently $204$ ks of \FUSE\ data available and a long
78 ks \HST-STIS E140M exposure for Mrk 205. Because of the low Complex C
column density in this sight line, this component cannot be detected in lines 
of \ion{O}{1}, \ion{Si}{2}, or \ion{Fe}{2} with \FUSE, since lines of
those species in the \FUSE\ bandpass are relatively weak compared to
their counterparts in the STIS E140M bandpass.  The only low-ion
absorption line in the \FUSE\ bandpass for which the Complex-C
component is detected is in the strong \ion{C}{2}$~\lambda1036.34$
line.  The Complex-C component is clearly detected in several
absorption lines in the STIS E140M bandpass, including the crucial
\ion{O}{1}~$\lambda1302.17$ line.  Several of the observed line profiles
are shown in Figure 3, along with the LDS profile of \ion{H}{1}
emission.  The LDS profile shows weak Complex-C emission at
$V_{\rm LSR} = -145$ \kms, along with strong H~I emission at 
$V_{\rm LSR} \approx 200$ \kms, from a {\it compact} high-velocity cloud 
labeled WW84 (Wakker \etal\ 2003) that is unrelated to Complex~C. 
Based on the observed absorption features in this sight line, we adopt 
an integration range of $-165<V_{LSR}<-125$ \kms\ for the Complex-C 
component.  Measured equivalent widths are shown in Table 3.

The Complex C component is detected in three \ion{Si}{2} and two \ion{C}{2}
lines.  We have fitted a curve of growth to those lines with doppler 
parameter $b=7.0^{+1.8}_{-1.1}$ \kms.  The resulting column 
densities are presented in Table 3.  Such a low doppler parameter suggests 
some amount of unresolved saturation in the profiles.  Because this
curve of growth is based on only a few lines, we hold some concern
for its accuracy when applied to other absorption lines beyond those
of \ion{Si}{2} and \ion{C}{2}, despite the small error bars on the
doppler parameter.  Since the \ion{O}{1} line is relatively weak, the
choice of doppler parameter has little influence on the resulting
\ion{O}{1} column density.  In fact, the value of $N$(\ion{O}{1})
calculated from the curve of growth differs from the value calculated
with the AOD method by only 0.07 dex.

We note that the Complex C component is not detected in any of the
ions ionized beyond \ion{Si}{3}.  Since the \ion{C}{4} absorption
feature near the Complex C integration range bears little resemblence
to other Complex C features, we suspect it may be spurious absorption
and unrelated to Complex C.  Limits on the column densities of the
high ions are presented in Table 3.  For sight lines that have easily
detectable low ions, this is the only Complex C sight line where
associated \ion{O}{6} is not detected.

\begin{deluxetable}{lrlcc}
\tablecolumns{5}
\tablewidth{0pc}
\tablecaption{SUMMARY OF MEASUREMENTS--MRK 205 SIGHTLINE\label{t3}}
\tablehead{
\colhead{} & \colhead{$\lambda$\tablenotemark{a}} & \colhead{} & \colhead{$W_{\lambda}$\tablenotemark{b}} & \colhead{log $N(X)$\tablenotemark{b}}  \\
\colhead{Species} & \colhead{(\AA)} & \colhead{$f$\tablenotemark{a}} & \colhead{(m\AA)} & \colhead{($N$ in cm$^{-2}$)} }
\startdata
\ion{H}{1}  & ...      \ \ & \ \ ...                     & ...       & $18.11^{+0.09}_{-0.11}(\pm 0.1)$\tablenotemark{c} \\
\ion{C}{2}  & 1036.337 \ \ & \ \ 0.118                   & $70\pm6$  & $14.48^{+0.15}_{-0.30}$ \\
...         & 1334.532 \ \ & \ \ 0.128                   & $106\pm3$ & ...                     \\
\ion{C}{4}  & 1548.204 \ \ & \ \ 0.190                   & $<25$     & $<12.84$                \\
\ion{N}{1}  & 1199.550 \ \ & \ \ 0.132                   & $<25$     & $<13.26$                \\
\ion{N}{5}  & 1238.821 \ \ & \ \ 0.156                   & $<16$     & $<12.85$                \\
\ion{O}{1}  & 1302.169 \ \ & \ \ 0.0480                  & $47\pm3$  & $13.99^{+0.06}_{-0.06}$ \\
\ion{O}{6}  & 1031.926 \ \  & \ \ 0.133                  & $<22$     & $<13.25$                \\
\ion{Al}{2} & 1670.787 \ \  & \ \ 1.740                  & $22\pm6$  & $11.77^{+0.12}_{-0.16}$ \\                   
\ion{Si}{2} & 1193.290 \ \  & \ \ 0.582                  & $50\pm12$ & $13.11^{+0.13}_{-0.09}$ \\
...         & 1260.422 \ \  & \ \ 1.18                   & $78\pm4$  & ...                     \\
...         & 1304.370 \ \  & \ \ 0.0863                 & $16\pm3$  & ...                     \\
\ion{Si}{3} & 1206.500 \ \  & \ \ 1.63                   & $>113$    & $>13.00$                \\
\ion{Si}{4} & 1393.760 \ \  & \ \ 0.513                  & $<23$     & $<12.42$                \\
\ion{S}{2}  & 1259.518 \ \  & \ \ 0.0166                 & $<14$     & $<13.82$                \\
\ion{Ar}{1} & 1048.220 \ \  & \ \ 0.263                  & $<17$     & $<12.89$                \\
\ion{Fe}{2} & 1608.451 \ \  & \ \ 0.0577                 & $<30$     & $<13.43$                \\
\enddata
\tablenotetext{a}{Wavelengths and oscillator strengths from Morton (2003).}
\tablenotetext{b}{Equivalent widths for Complex C integrated over velocity 
range $-165$ to $-125$ \kms.  Error bars on measured column densities are 
$1 \sigma$; upper limits are 3$\sigma$. Column densities are calculated 
from curve of growth with doppler parameter, 
$b=7.0^{+1.8}_{-1.1}$ \kms, except \ion{O}{6}, \ion{N}{5}, \ion{C}{4}, 
\ion{Si}{4}, and \ion{Si}{3} where we used the AOD method.} 
\tablenotetext{c}{The \NHI\ column density includes a systematic error 
(in parentheses) of 0.1 dex due to beam-size mismatch between \ion{H}{1} 
21-cm data and quasar absorption spectra.}
\end{deluxetable}

\begin{figure*}
\figurenum{3}
\epsscale{1}
\plotone{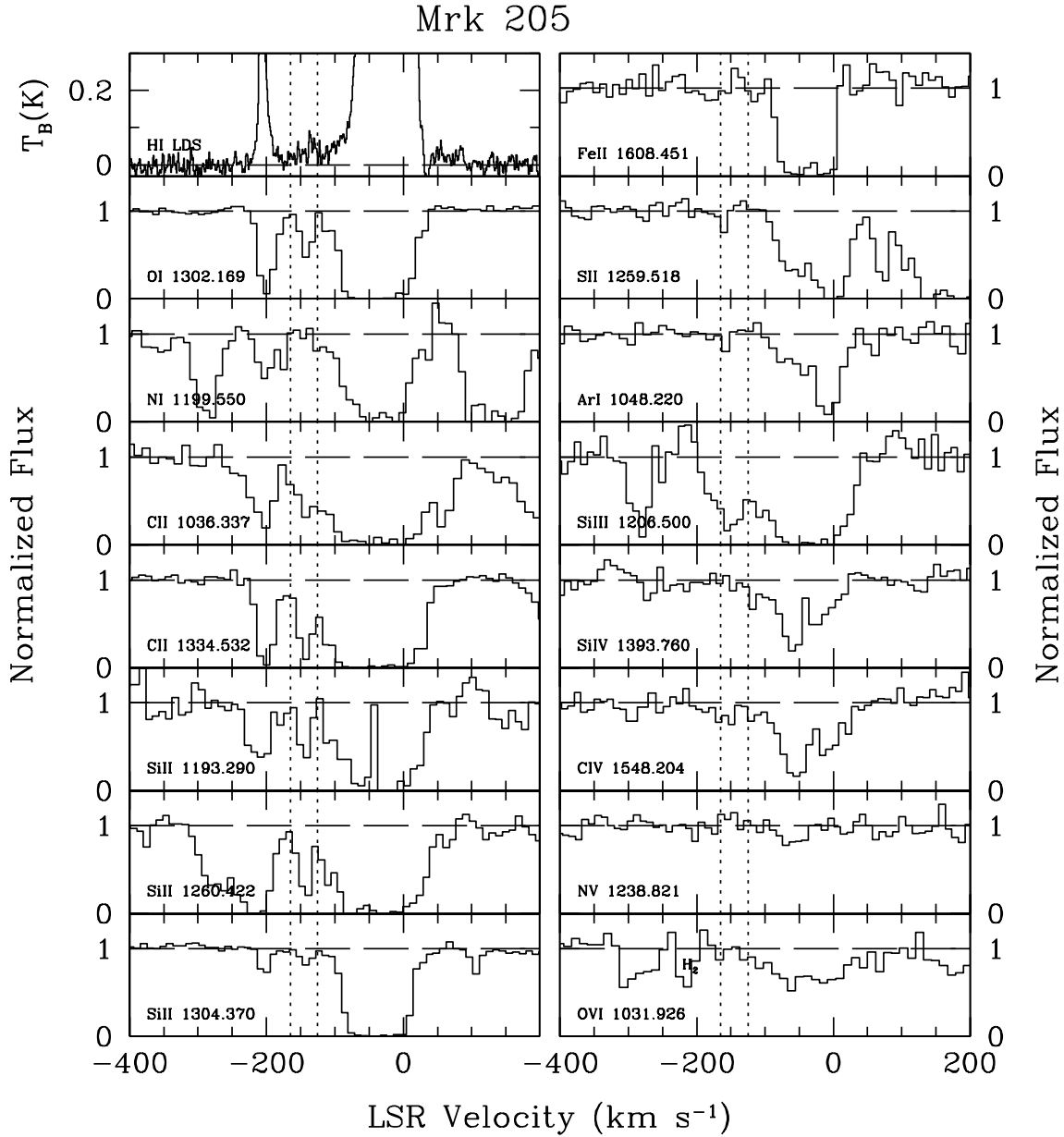}
\caption{A sample of normalized absorption profiles from \FUSE\ and \HST\ 
STIS E140M data for Mrk 205, along with the LDS profile of 
\ion{H}{1} emission.  The vertical dashed lines indicate the $-165$ to 
$-125$ \kms\ range of integration for measurements of $W_{\lambda}$.}
\end{figure*} 

\begin{deluxetable}{lrlcc}
\tablecolumns{5}
\tablewidth{0pc}
\tablecaption{SUMMARY OF MEASUREMENTS--MRK 279 SIGHTLINE\label{t4}}
\tablehead{
\colhead{} & \colhead{$\lambda$\tablenotemark{a}} & \colhead{} & \colhead{$W_{\lambda}$\tablenotemark{b}} & \colhead{log $N(X)$\tablenotemark{b}}  \\
\colhead{Species} & \colhead{(\AA)} & \colhead{$f$\tablenotemark{a}} & \colhead{(m\AA)} & \colhead{($N$ in cm$^{-2}$)} }
\startdata
\ion{H}{1}  & ...      \ \ & \ \ ...                     & ...       & $19.27^{+0.08}_{-0.11}(\pm 0.1)$\tablenotemark{c} \\
\ion{C}{4}  & 1548.204 \ \ & \ \ 0.190                   & $73\pm6$  & $13.30^{+0.04}_{-0.04}$ \\
\ion{N}{1}  & 1199.550 \ \ & \ \ 0.132                   & $26\pm6$  & $13.22^{+0.10}_{-0.12}$ \\
\ion{N}{2}  & 1083.994 \ \ & \ \ 0.111                   & $150\pm7$ & $14.45^{+0.11}_{-0.10}$ \\
\ion{N}{5}  & 1238.821 \ \ & \ \ 0.156                   & $<14$     & $<12.84$                \\
\ion{O}{1}  & 929.517 \ \   & \ \ 0.00229                & $19\pm6$  & $14.98^{+0.10}_{-0.06}$ \\
...         & 936.630 \ \   & \ \ 0.00365                & $28\pm6$ & ...                      \\
...         & 948.686 \ \   & \ \ 0.00631                & $41\pm7$ & ...                      \\
...         & 971.738 \ \   & \ \ 0.0116                 & $75\pm6$ & ...                      \\
...         & 976.448 \ \   & \ \ 0.00331                & $24\pm6$ & ...                      \\
...         & 1039.230 \ \  & \ \ 0.00907                & $68\pm3$ & ...                      \\
...         & 1302.169 \ \  & \ \ 0.0480                 & $208\pm4$ & ...                     \\
\ion{O}{6}  & 1031.926 \ \  & \ \ 0.133                  & $52\pm4$  & $13.66^{+0.04}_{-0.04}$ \\
\ion{Al}{2} & 1670.787 \ \  & \ \ 1.740                  & $208\pm9$ & $12.96^{+0.09}_{-0.07}$ \\                   
\ion{Si}{2} & 1020.699 \ \  & \ \ 0.0168                 & $22\pm4$  & $14.14^{+0.06}_{-0.08}$ \\
...         & 1190.416 \ \  & \ \ 0.292                  & $198\pm6$ & ...                     \\
...         & 1193.290 \ \  & \ \ 0.582                  & $244\pm6$ & ...                     \\
...         & 1304.370 \ \  & \ \ 0.0863                 & $114\pm4$ & ...                     \\
...         & 1526.707 \ \  & \ \ 0.133                  & $183\pm6$ & ...                     \\
\ion{Si}{4} & 1393.760 \ \  & \ \ 0.513                  & $45\pm6$  & $12.73^{+0.05}_{-0.05}$ \\
\ion{P}{2}  & 963.801  \ \  & \ \ 1.460                  & $<17$     & $<12.18$                \\
\ion{S}{2}  & 1259.518 \ \  & \ \ 0.0166                 & $11\pm3$  & $13.69^{+0.11}_{-0.15}$ \\
\ion{Ar}{1} & 1048.220 \ \  & \ \ 0.263                  & $<12$     & $<12.69$                \\
\ion{Fe}{2} & 1096.877 \ \  & \ \ 0.0327                 & $22\pm4$  & $13.85^{+0.11}_{-0.10}$ \\
...         & 1121.975 \ \  & \ \ 0.0290                 & $17\pm4$ & ...                      \\
...         & 1125.448 \ \  & \ \ 0.0156                 & $15\pm5$  & ...                     \\
...         & 1144.938 \ \  & \ \ 0.0830                 & $55\pm5$  & ...                     \\
...         & 1608.451 \ \  & \ \ 0.0577                 & $80\pm6$  & ...                     \\
\enddata
\tablenotetext{a}{Wavelengths and oscillator strengths from Morton (2003).}
\tablenotetext{b}{Equivalent widths for Complex C are integrated over velocity 
range $-200$ to $-120$ \kms.  Error bars on measured column densities are 
$1 \sigma$; upper limits are 3$\sigma$. Column densities are calculated from 
curve of growth with doppler parameter, 
$b=18.0^{+2.0}_{-1.5}$ \kms, except for \ion{O}{6}, \ion{N}{5}, \ion{C}{4}, 
and \ion{Si}{4}, where we used AOD method.} 
\tablenotetext{c}{The \NHI\ column density includes a systematic error 
(in parentheses) of 0.1 dex because of beam-size mismatch between \ion{H}{1} 
21-cm data and quasar absorption spectra.}
\end{deluxetable}

\begin{figure*}
\figurenum{4}
\epsscale{1}
\plotone{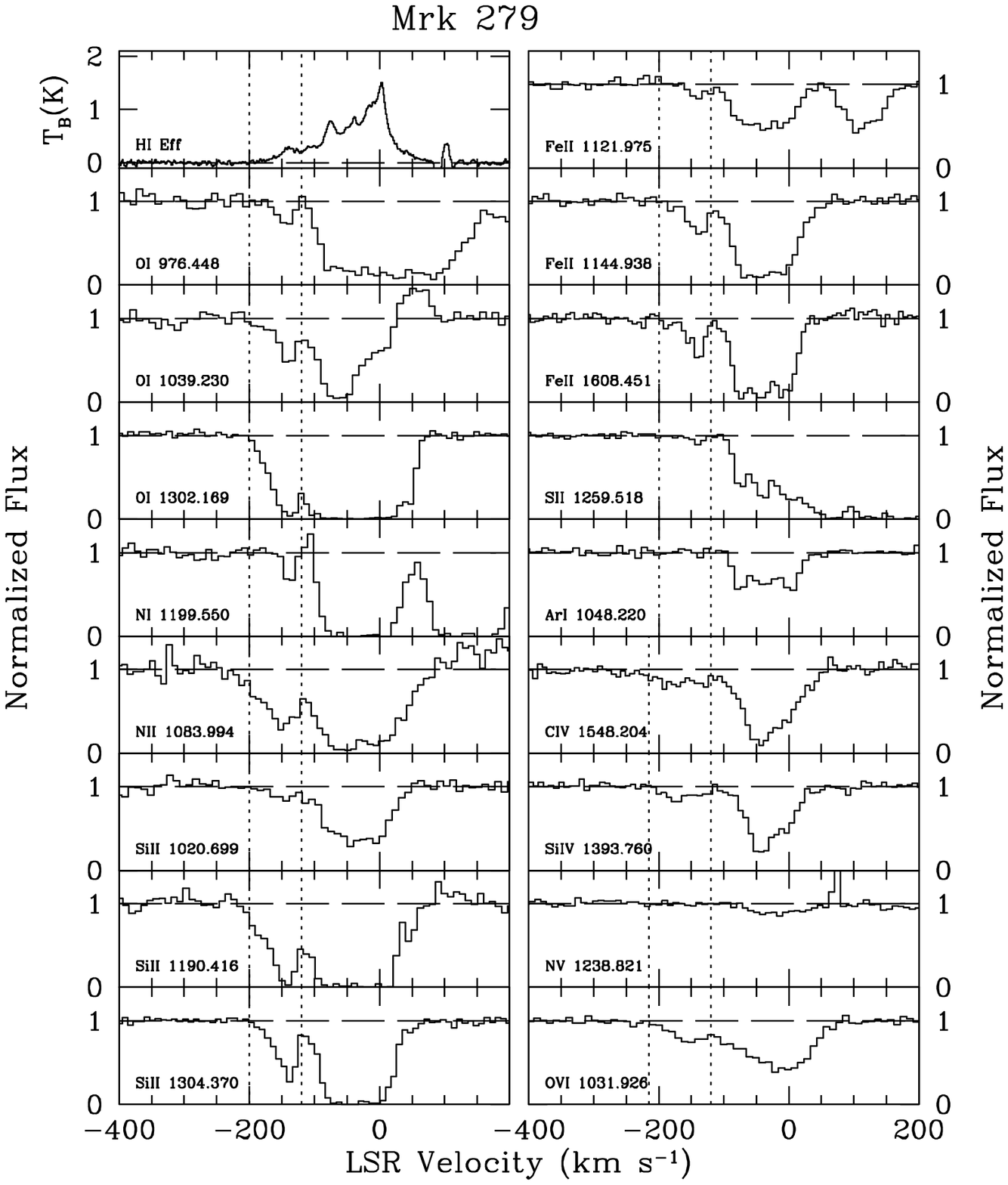}
\caption{A sample of normalized absorption profiles from \FUSE\ and \HST\ 
STIS E140M data for Mrk 279, along with the Effelsberg profile of 
\ion{H}{1} emission.  The vertical dashed lines indicate the $-200$ to 
$-120$ \kms\ range of integration for measurements of $W_{\lambda}$, 
except for the high ions, \ion{C}{4}, \ion{Si}{4}, \ion{N}{5}, 
and \ion{O}{6}, where 
measurements extend to as low as $-215$ \kms.}
\end{figure*} 

\subsection{Mrk 279}

Since CSG03, 137~ks of additional \FUSE\ data have been taken for
this sight line, and extensive \HST-STIS E140M exposures have
recently been obtained.  The \FUSE\ and
\HST\ spectra were observed simultaneously in both 2002 and 2003.
The 2002 spectra indicate that the quasar flux is lower by at least a
factor of 5 than the original 1999/2000 \FUSE\ exposure and the
2003 spectra.  We therefore do not include the 2002 data in our final
Mrk 279 spectrum.  

\begin{figure*}
\figurenum{5}
\epsscale{1}
\plotone{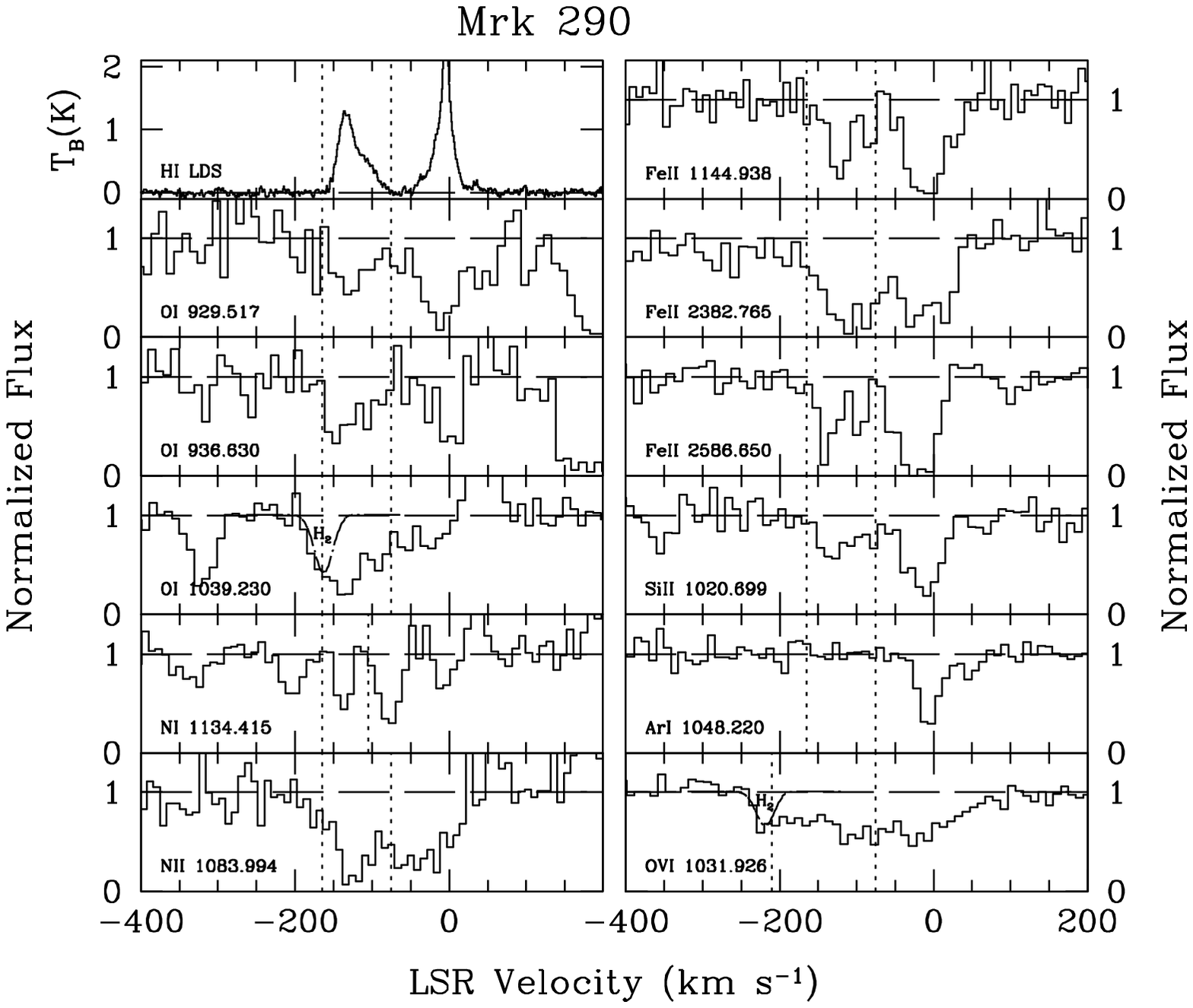}
\caption{A sample of normalized absorption profiles from \FUSE\ and \HST\ 
STIS E230M data for Mrk 290, along with the LDS profile of \ion{H}{1} 
emission.  The vertical dashed lines indicate the $-165$ to $-75$ \kms\ 
range of integration for measurements of $W_{\lambda}$, except for \ion{O}{6}, 
where measurements extend to a minimum velocity of $-210$ \kms.  The 
\ion{O}{1} and \ion{O}{6} profiles include a fit to \Htwo\ contamination.}
\end{figure*} 

The Mrk 279 \FUSE\ and STIS data have the highest signal to
noise among the Complex C sight lines.  Samples of the analyzed
line profiles are shown in Figure 4, along with the Effelsberg profile
of \ion{H}{1} emission.  The component structure along this sight line
is complicated; Wakker \etal\ (2003) identified eight components to
the \ion{H}{1} profile, two of which are attributed to
Complex C.  CSG03 adopted the approach of integrating over both HVC
components, even though contamination from higher-metallicity components 
at lower velocity was a possibility.  Here, we take the
approach of integrating over only the highest-velocity Complex C
component, $-200<V_{LSR}<-120$ \kms, so as to be reasonably sure 
that no contamination from non-Complex C absorption occurs.  The high-ions,
particularly \ion{C}{4} and \ion{O}{6}, extend more blueward than the
lower ions.  Therefore, we extend the integration range for measurements of 
the highly ionized component out to $V_{LSR}=-215$ \kms.  The weaker 
lines of the \ion{C}{4} and \ion{Si}{4} doublets were not detected, 
whereas the stronger lines (C~IV~$\lambda1548$ and Si~IV~$\lambda 1393$) 
were seen at optical depths $\tau \approx 0.2$ in 7 \kms\ resolution bins.  
We believe the weaker lines were on the margin of detectability
(at $\sim2 \sigma$ level) at S/N $\approx 19-26$ per resolution element. 

Equivalent widths for measurements of the highest-velocity Complex C
component are shown in Table 4.  We have detected the Complex C
component in seven \ion{O}{1}, five \ion{Si}{2}, and five \ion{Fe}{2} lines.
>From the equivalent width measurements of these lines, we have
determined a best fit curve of growth with doppler parameter,
$b=18.0^{+2.0}_{-1.5}$ \kms.  Resulting column densities of the
various ion species are shown in Table 4.  The \ion{H}{1} column
density was determined by direct integration of the Effelsberg profile
over the specified velocity range.

\subsection{Mrk 290}

\begin{deluxetable}{lrlcc}
\tablecolumns{5}
\tablewidth{0pc}
\tablecaption{SUMMARY OF MEASUREMENTS--MRK 290 SIGHTLINE\label{t5}}
\tablehead{
\colhead{} & \colhead{$\lambda$\tablenotemark{a}} & \colhead{} & \colhead{$W_{\lambda}$\tablenotemark{b}} & \colhead{log $N(X)$\tablenotemark{b}}  \\
\colhead{Species} & \colhead{(\AA)} & \colhead{$f$\tablenotemark{a}} & \colhead{(m\AA)} & \colhead{($N$ in cm$^{-2}$)} }
\startdata
\ion{H}{1}  & ...      \ \  & \ \ ...                     & ...        & $20.12^{+0.02}_{-0.02}(\pm 0.1)\tablenotemark{c}$ \\
\ion{N}{1}  & 1134.415 \ \  & \ \ 0.0287                  & $46\pm14$  & $14.23^{+0.17}_{-0.20}$ \\
\ion{N}{2}  & 1083.994 \ \  & \ \ 0.115                   & $190\pm28$ & $14.96^{+1.04}_{-0.48}$ \\
\ion{N}{5}  & 1238.821 \ \  & \ \ 0.156                   & $<65$      & $<13.49$                \\
\ion{O}{1}  & 929.517  \ \  & \ \ 0.00229                 & $77\pm18$  & $15.75^{+0.29}_{-0.15}$ \\
...         & 936.630  \ \  & \ \ 0.00365                 & $105\pm17$ & ...                     \\
...         & 950.885  \ \  & \ \ 0.00158                 & $63\pm20$  & ...                     \\ 
...         & 971.738  \ \  & \ \ 0.0116                  & $153\pm20$ & ...                     \\
...         & 976.448  \ \  & \ \ 0.00331                 & $69\pm19$  & ...                     \\
...         & 1039.230 \ \  & \ \ 0.00907                 & $142\pm10$ & ...                     \\
\ion{O}{6}  & 1031.926 \ \  & \ \ 0.133                   & $163\pm15$ & $14.23^{+0.04}_{-0.04}$ \\
\ion{Si}{2} & 1020.699 \ \  & \ \ 0.0168                  & $83\pm12$  & $14.93^{+0.18}_{-0.13}$ \\
\ion{P}{2}  & 963.801  \ \  & \ \ 1.460                   & $<60$      & $<12.82$                \\
\ion{S}{2}  & 1259.518 \ \  & \ \ 0.0166                  & $36\pm12$  & $14.24^{+0.16}_{-0.20}$ \\
\ion{Ar}{1} & 1048.220 \ \  & \ \ 0.263                   & $<32$      & $<13.15$                \\
\ion{Mn}{2} & 2576.877 \ \  & \ \ 0.361                   & $<99$      & $<12.74$                \\
\ion{Fe}{2} & 1121.975 \ \  & \ \ 0.0290                  & $68\pm13$  & $14.41^{+0.23}_{-0.21}$ \\
...         & 1144.938 \ \  & \ \ 0.0830                  & $138\pm18$ & ...                     \\
...         & 2344.214 \ \  & \ \ 0.114                   & $361\pm36$ & ...                     \\
...         & 2374.461 \ \  & \ \ 0.0313                  & $199\pm48$ & ...                     \\
...         & 2382.765 \ \  & \ \ 0.320                   & $506\pm28$ & ...                     \\
...         & 2586.650 \ \  & \ \ 0.0691                  & $317\pm31$ & ...                     \\
...         & 2600.173 \ \  & \ \ 0.239                   & $474\pm27$ & ...                     \\
\enddata
\tablenotetext{a}{Wavelengths and oscillator strengths from Morton (2003).}
\tablenotetext{b}{Equivalent widths for Complex C are integrated over velocity 
range $-165$ to $-75$ \kms.   Error bars on measured column densities are
$1 \sigma$; upper limits are 3$\sigma$.  
Column densities are calculated from curve of growth with doppler 
parameter, $b=15.7^{+3.6}_{-2.8}$ \kms, except for \ion{O}{6} and \ion{N}{5}, 
where we used the AOD method.} 
\tablenotetext{c}{The \NHI\ column density includes a systematic error (in 
parentheses) of 0.1 dex, arising from beam-size mismatch between \ion{H}{1} 
21-cm data and quasar absorption spectra.}
\end{deluxetable}

\begin{deluxetable*}{lrlcccccc}
\tablecolumns{7}
\tablewidth{0pc}
\tablecaption{SUMMARY OF MEASUREMENTS--MRK 876 SIGHTLINE\label{t6}}
\tablehead{
\colhead{} & \colhead{} & \colhead{} & \multicolumn{2}{c}{Component 1} & \multicolumn{2}{c}{Component 2} \\
\colhead{} & \colhead{} & \colhead{} & \multicolumn{2}{c}{$V_{LSR}=-135$ km s$^{-1}$} & \multicolumn{2}{c}{$V_{LSR}=-175$ km s$^{-1}$} \\
\colhead{} & \colhead{$\lambda$\tablenotemark{a}} & \colhead{} & \colhead{$W_{\lambda}$\tablenotemark{b}} & \colhead{log $N(X)$\tablenotemark{b}} & \colhead{$W_{\lambda}$\tablenotemark{b}} & \colhead{log $N(X)$\tablenotemark{b}}  \\
\colhead{Species} & \colhead{(\AA)} & \colhead{$f$\tablenotemark{a}} & \colhead{(m\AA)} & \colhead{($N$ in cm$^{-2}$)} & \colhead{(m\AA)} & \colhead{($N$ in cm$^{-2}$)} }
\startdata
\ion{H}{1}  & ...      \ \  & \ \ ...                     & ...        & $19.30^{+0.03}_{-0.04}(\pm 0.1)$\tablenotemark{c} & ...        & $18.72^{+0.12}_{-0.17}(\pm 0.1)$\tablenotemark{c} \\
\ion{N}{1}  & 1199.550 \ \  & \ \ 0.132                   & $96\pm13$  & $13.90^{+0.12}_{-0.10}$ & $<35$      & $<13.37$                \\
\ion{N}{2}  & 1083.994 \ \  & \ \ 0.111                   & $188\pm16$ & $14.75^{+0.35}_{-0.22}$ & $53\pm15$  & $13.76^{+0.18}_{-0.19}$ \\
\ion{O}{1}  & 936.630  \ \  & \ \ 0.00365                 & $51\pm15$  & $15.26^{+0.17}_{-0.14}$ & $<40$      & $14.79^{+0.18}_{-0.11}$  \\
...         & 948.686  \ \  & \ \ 0.00631                 & ...        & ...                     & $35\pm11$  & ...                      \\
...         & 971.738  \ \  & \ \ 0.0116                  & ...        & ...                     & $56\pm12$  & ...                      \\
...         & 976.448  \ \  & \ \ 0.00331                 & $56\pm15$  & ...                     & $<38$      & ...                      \\
...         & 1039.230 \ \  & \ \ 0.00907                 & $92\pm10$  & ...                     & ...        & ...                      \\
...         & 1302.169 \ \  & \ \ 0.0480                  & $253\pm7$  & ...                     & $169\pm6$  & ...                      \\
\ion{Al}{2} & 1670.787 \ \  & \ \ 1.740                   & $301\pm16$ & $13.43^{+0.31}_{-0.19}$ & $103\pm15$ & $12.51^{+0.13}_{-0.11}$  \\
\ion{Si}{2} & 1020.699 \ \  & \ \ 0.0168                  & $42\pm5$   & $14.56^{+0.11}_{-0.09}$ & $<14$      & $13.85^{+0.09}_{-0.09}$  \\
...         & 1190.416 \ \  & \ \ 0.292                   & $245\pm9$  & ...                     & $140\pm8$  & ...                      \\
...         & 1193.290 \ \  & \ \ 0.582                   & $251\pm10$ & ...                     & $173\pm8$  & ...                      \\
...         & 1304.370 \ \  & \ \ 0.0863                  & $202\pm7$  & ...                     & $69\pm7$   & ...                      \\
...         & 1526.707 \ \  & \ \ 0.133                   & $261\pm10$ & ...                     & $116\pm9$  & ...                      \\
\ion{S}{2}  & 1253.805 \ \  & \ \ 0.0109                  & $<31$      & $<14.34$                & $<26$      & $<14.26$                 \\
\ion{Ar}{1} & 1048.220 \ \  & \ \ 0.263                   & $<15$      & $<12.79$                & $<16$      & $<12.83$                 \\
\ion{Fe}{2} & 1063.176 \ \  & \ \ 0.0547                  & ...        & $14.36^{+0.07}_{-0.07}$ & $23\pm5$   & $13.72^{+0.15}_{-0.13}$  \\
...         & 1121.975 \ \  & \ \ 0.0290                  & $49\pm5$   & ...                     & $<12$      & ...                      \\
...         & 1125.448 \ \  & \ \ 0.0156                  & $37\pm5$   & ...                     & $<12$      & ...                      \\
...         & 1143.226 \ \  & \ \ 0.0192                  & $39\pm5$   & ...                     & $<14$      & ...                      \\
...         & 1144.938 \ \  & \ \ 0.0830                  & $131\pm5$  & ...                     & $41\pm5$   & ...                      \\
...         & 1608.451 \ \  & \ \ 0.0577                  & $187\pm17$ & ...                     & $72\pm15$  & ...                      \\
\enddata
\tablenotetext{a}{Wavelengths and oscillator strengths from Morton (2003).}
\tablenotetext{b}{Equivalent widths for Complex C are integrated over velocity 
range $-155$ to $-85$ \kms\ for Component 1 and $-210$ to $-155$ \kms\ for 
Component 2.   Error bars on measured column densities are $1 \sigma$; upper 
limits are 3$\sigma$.  Column densities are calculated from 
curves of growth with doppler parameters, $b=17.8^{+2.0}_{-1.8}$ \kms\ for 
Component 1 and $b=15.4^{+3.5}_{-2.5}$ \kms\ for Component 2, except for 
\ion{O}{6} and \ion{N}{5}, where we used the AOD method.} 
\tablenotetext{c}{The \NHI\ column density includes a systematic error (in 
parentheses) of 0.1 dex, arising from beam-size mismatch between \ion{H}{1} 
21-cm data and quasar absorption spectra.}
\end{deluxetable*}

\begin{figure*}
\figurenum{6}
\epsscale{1}
\plotone{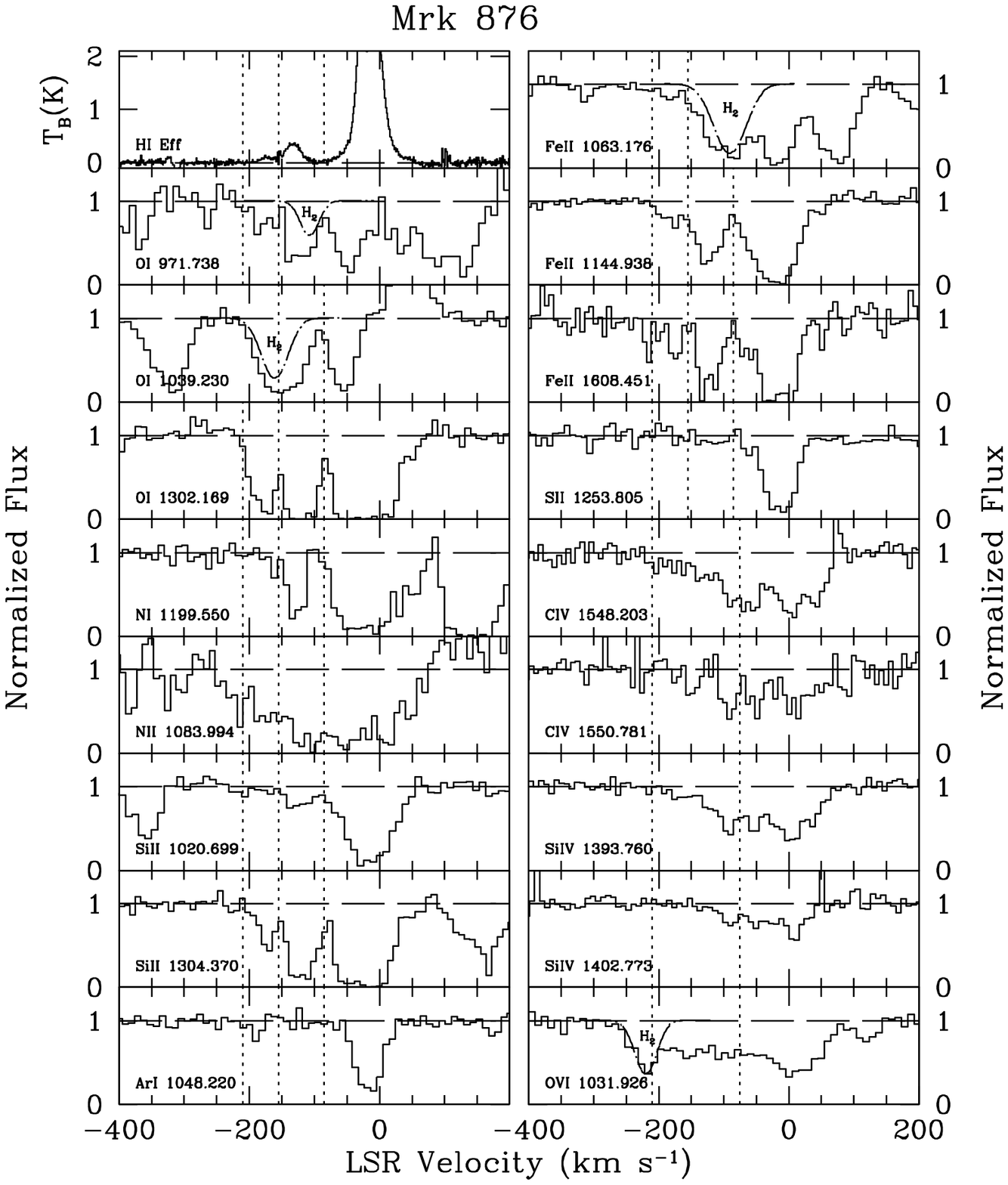}
\caption{A sample of normalized absorption profiles from \FUSE\ and \HST\ 
STIS E140M data for Mrk 876, along with the Effelsberg profile of \ion{H}{1} 
emission.  The vertical dashed lines indicate the $-155$ to $-85$ \kms\
and $-210$ to $-155$ \kms\ ranges of integration for profiles where we 
make measurements of $W_{\lambda}$ for Components 1 and 2 of Complex C. For 
the high ions, \ion{C}{4}, \ion{Si}{4}, \ion{N}{5}, and \ion{O}{6}, our 
measurements are made over the range $-210$ to $-75$ \kms.  Also 
included are fits to contaminating \Htwo\ absorption.}
\end{figure*}

\begin{deluxetable}{lrlcc}
\tablecolumns{5}
\tablewidth{0pc}
\tablecaption{COMPLEX C HIGHLY IONIZED COMPONENT--MRK 876 SIGHTLINE\tablenotemark{a} 
\label{t7} }
\tablehead{
\colhead{} & \colhead{$\lambda$\tablenotemark{a}} & \colhead{} & \colhead{$W_{\lambda}$\tablenotemark{b}} & \colhead{log $N(X)$\tablenotemark{c}}  \\
\colhead{Species} & \colhead{(\AA)} & \colhead{$f$\tablenotemark{b}} & \colhead{(m\AA)} & \colhead{($N$ in cm$^{-2}$)} }
\startdata
\ion{C}{4}  & 1548.204 \ \ & \ \ 0.190                   & $199\pm19$ & $13.80^{+0.04}_{-0.03}$ \\
...         & 1550.781 \ \ & \ \ 0.0948                  & $96\pm20$  & $13.81^{+0.06}_{-0.08}$ \\
\ion{Si}{4} & 1393.760 \ \ & \ \ 0.513                   & $134\pm9$  & $13.28^{+0.03}_{-0.02}$ \\
...         & 1402.773 \ \ & \ \ 0.254                   & $41\pm9$   & $13.01^{+0.09}_{-0.10}$ \\
\ion{N}{5}  & 1238.821 \ \ & \ \ 0.156                   & $<43$      & $<13.32$                \\
\ion{O}{6}  & 1031.926 \ \ & \ \ 0.133                   & $153\pm9$  & $14.20^{+0.02}_{-0.02}$ \\
\enddata
\tablenotetext{a}{Data for the high ions are presented separately since they 
have a distinctly different velocity range and profile than the low ions 
presented in Table 4. }
\tablenotetext{b}{Wavelengths and oscillator strengths from Morton (2003).
 Equivalent widths of the highly ionized component are measured over 
 velocity range $-210$ to $-75$ \kms. }
\tablenotetext{c}{ Error bars on measured column densities are $1 \sigma$;
upper limits are 3$\sigma$. Column densities are measured with the AOD method.}  
\end{deluxetable}

The Mrk 290 sight line was the first Complex C target investigated with
UV spectra.  Using Goddard High-Resolution Spectrograph (GHRS) data for 
Mrk 290, Wakker \etal\ (1999) determined
a Complex C metallicity of 0.1 $Z_{\sun}$ based on the 
\ion{S}{2}/\ion{H}{1} abundance.  The CSG03 analysis used a brief \FUSE\ 
exposure to measure upper limits on ion abundances.  Since CSG03, a substantial
\FUSE\ exposure has been obtained to supplement the previously available GHRS 
data.  In addition, an archival STIS E230M dataset is available, which 
covers strong \ion{Fe}{2} lines in the 2300-2600 \AA\ range.

Figure 5 shows a sample of the analyzed line profiles, along with the
LDS profile of
\ion{H}{1} emission.  Based on the various line profiles, we adopt an
integration range of $-165<V_{LSR}<-75$ \kms, 
except for \ion{O}{6} which extends blueward to $-210$ \kms.  We 
see evidence for a two-component structure in the Complex C absorption 
for several absorption lines, most notably the \ion{Fe}{2} lines in the
higher signal-to-noise STIS data.  However, the primary goal of this
work is to determine \ion{O}{1} abundances using a technique that
relies heavily on weaker \ion{O}{1} lines in the SiC channels.  The SiC 
data for Mrk 290 have low signal to noise, and multi-component structure
is not apparent for most \ion{O}{1} lines, although high-velocity
absorption is clearly present.  As a result, we consider the entire
Complex C velocity range as a single feature.  Because \Htwo\ contamination 
is a problem for several Complex C absorption features in the \FUSE\ 
bandpass, we have modeled the expected profiles of contaminant
\Htwo\ lines, shown also in Figure 5.

Equivalent width measurements of Complex C absorption in this sight
line are shown in Table 5.  We also include
\ion{S}{2}~$\lambda1259.518$ and \ion{N}{5}~$\lambda1238.821$ results
from the GHRS data.  Using results from the 6 \ion{O}{1} and 7
\ion{Fe}{2} line measurements, we determine a best-fit curve of growth
with doppler parameter, $b=13.4^{+4.3}_{-2.9}$ \kms.  Resulting column
densities are listed in Table 5.

\subsection{Mrk 876}

The Complex C Mrk 876 sight line was investigated both by Murphy \etal\ 
(2000) and CSG03.  These studies were unable to make measurements of 
\ion{O}{1} owing to strong \Htwo\ contamination and poor data quality in 
the SiC channels.  Since those publications, 80~ks of additional \FUSE\ 
data and a high-quality \HST-STIS E140M dataset have become available.  

A sample of the analyzed line profiles are shown in Figure 6, along
with the Effelsberg profile of \ion{H}{1} emission.  Both the
\ion{H}{1} analysis by Wakker et al. (2003) and the line profiles
presented here indicate a two-component structure for Complex C in
this direction.  Based on these profiles, we adopt integration ranges
of $-155<V_{LSR}<-85$ \kms\ and $-210< V_{LSR} < -155$ \kms, 
which we identify as Component 1 (centered at $V_{LSR}\approx-135$ \kms) 
and Component 2 (centered at $V_{LSR}\approx-175$ \kms), respectively.
This sight line pierces a particularly thick column of Galactic \Htwo,
N(\Htwo) $\approx 4.4 \times 10^{16}$ \cd\ (Gillmon \etal\ 2006).    
Fits to contaminant \Htwo\ lines are shown in several of the profiles.  

Table 6 shows equivalent width measurements for Components 1 and 2 of lines 
in the \FUSE\ and \HST-STIS E140M bandpass.  Multiple lines of \ion{O}{1}, 
\ion{Si}{2}, and \ion{Fe}{2} are detected for each component.  To those 
measurements, we fitted curves of growth to Components 1 and 2 of 
$b=17.8^{+2.0}_{-1.8}$ \kms and $b=15.4^{+3.5}_{-2.5}$ \kms, 
respectively.  Table 6 also shows the resulting column densities calculated 
from the curves of growth.  We calculate \NHI\ for each component by direct 
integration of the Effelsberg \ion{H}{1}\ profile.  
 
\begin{deluxetable}{lrlcc}
\tablecolumns{5}
\tablewidth{0pc}
\tablecaption{SUMMARY OF MEASUREMENTS--PG 1626+554 SIGHTLINE\label{t8}}
\tablehead{
\colhead{} & \colhead{$\lambda$\tablenotemark{a}} & \colhead{} & \colhead{$W_{\lambda}$\tablenotemark{b}} & \colhead{log $N(X)$\tablenotemark{b}}  \\
\colhead{Species} & \colhead{(\AA)} & \colhead{$f$\tablenotemark{a}} & \colhead{(m\AA)} & \colhead{($N$ in cm$^{-2}$)} }
\startdata
\ion{H}{1}  & ...      \ \  & \ \ ...                     & ...        & $19.43^{+0.02}_{-0.02}(\pm 0.1)$\tablenotemark{c} \\
\ion{N}{1}  & 1134.165 \ \  & \ \ 0.0146                  & $<50$      & $<14.49$                \\
\ion{O}{1}  & 1039.230 \ \  & \ \ 0.00907                 & $107\pm12$ & $[15.30^{+0.08}_{-0.08},15.68^{+0.22}_{-0.19}]$ \\
\ion{O}{6}  & 1031.926 \ \  & \ \ 0.133                   & $119\pm16$ & $14.10^{+0.05}_{-0.06}$ \\
\ion{Si}{2} & 1020.699 \ \  & \ \ 0.0168                  & $46\pm15$  & $14.56^{+0.11}_{-0.16}$ \\
\ion{Ar}{1} & 1048.220 \ \  & \ \ 0.263                   & $<44$      & $<13.25$                \\
\ion{Fe}{2} & 1144.938 \ \  & \ \ 0.0830                  & $102\pm16$ & $[14.19^{+0.10}_{-0.11},14.45^{+0.22}_{-0.19}]$ \\
\enddata
\tablenotetext{a}{Wavelengths and oscillator strengths from Morton (2003).}
\tablenotetext{b}{Equivalent widths for Complex C are integrated over 
velocity range $-155$ to $-75$ \kms.  Error bars on measured column 
densities are $1 \sigma$; upper limits are 3$\sigma$.  
Ranges of column densities for \ion{O}{1}\ and \ion{Fe}{2}\
are calculated from curves of growth with doppler 
parameters, $b=10.2$ \kms\ and $b=18.0$ \kms\ (in brackets). 
The AOD method is used for all other species.} 
\tablenotetext{c}{The \NHI\ column density includes a systematic error 
(in parentheses) of 0.1 dex because of beam-size mismatch between \ion{H}{1} 
21-cm data and quasar absorption spectra.}
\end{deluxetable}

\begin{figure*}
\figurenum{7}
\epsscale{1}
\plotone{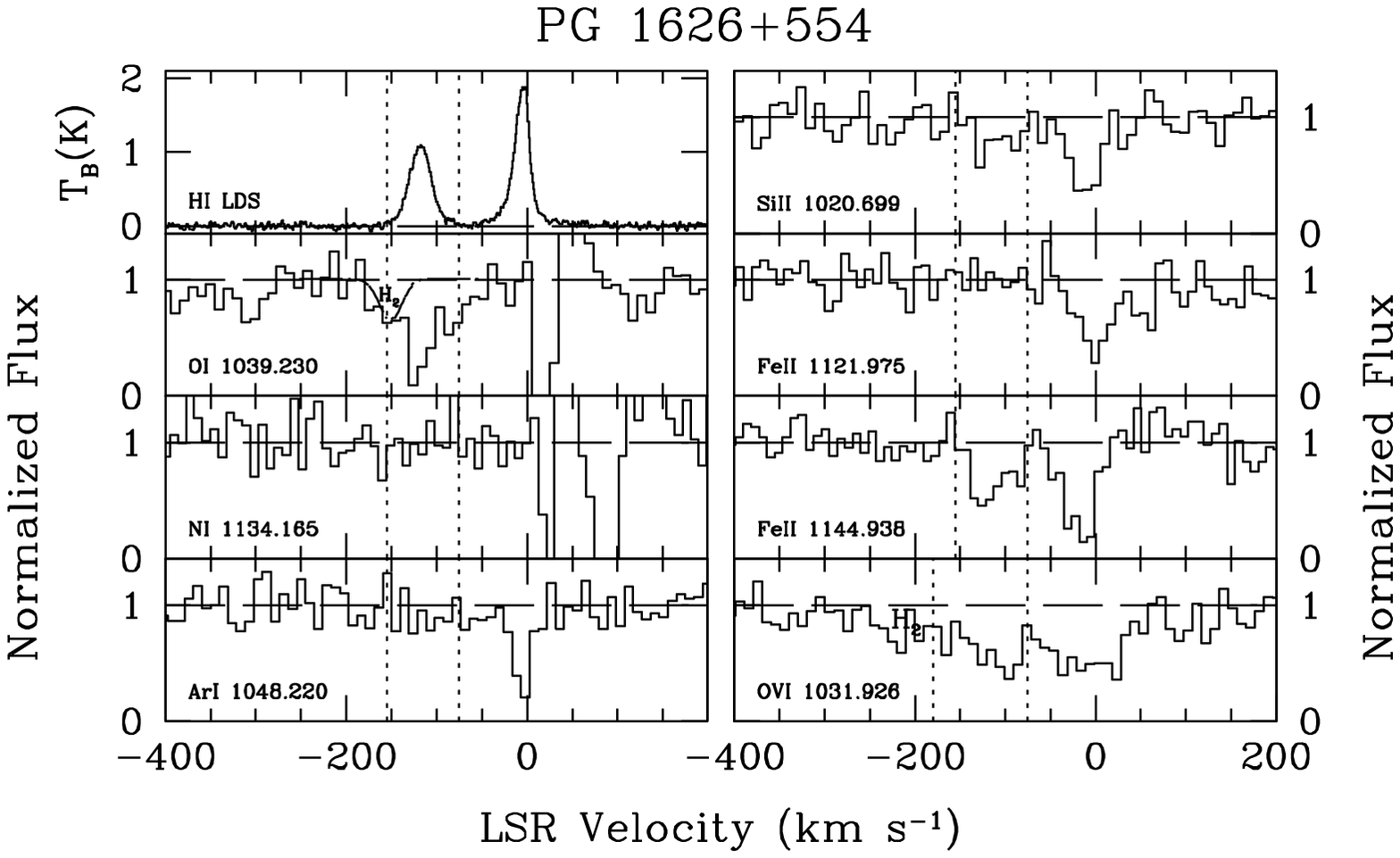}
\caption{A sample of normalized absorption profiles from \FUSE\ data for 
PG~1626+554, along with the LDS profile of \ion{H}{1} emission.  The vertical 
dashed lines indicate the $-155$ to $-75$ \kms\ range of integration 
for measurements of $W_{\lambda}$, except for \ion{O}{6}, where measurements 
extend to a minimum velocity of $-180$ \kms.  The \ion{O}{1} profile 
includes a fit to \Htwo\ contamination.}
\end{figure*} 

The high ions \ion{C}{4} and \ion{Si}{4} show an unusual
velocity structure, which is offset to lower velocities from the low
ions.  The high-ion profiles show peak absorption at 
$V_{LSR}\approx-95$ \kms, well redward of the low ions.  
Additionally, we detect little evidence of the two-component structure 
seen in the low
ions, except possibly in the case of \ion{C}{4}~$\lambda1550.781$.  We
adopt an integration range for the high-ions of $-210<V_{LSR}<-75$
\kms\ and analyze those lines separately from Components 1 and 2.
High-ion column densities of this component, calculated with the
apparent optical depth method, are listed in Table 7.

\subsection{PG 1626+554}

At the time of publication of CSG03, only 8~ks of \FUSE\ data were
available for this sight line.  An additional 90~ks of data were
recently taken using the MDRS slit.  Although the target is not
present in the new SiC data, the current dataset does allow
measurement of the strongest absorption lines in the 1000-1150 \AA\
range.

Several of the observed line profiles are shown in Figure 7, along
with the LDS profile of \ion{H}{1} emission.  We are able to make
$>3\sigma$ measurements for only four absorption lines, including
\ion{O}{1} $\lambda1039.230$, \ion{O}{6} $\lambda1031.926$,
\ion{Si}{2} $\lambda1020.699$, and \ion{Fe}{2} $\lambda1144.938$.  Based
on these profiles, we adopt an integration range of $-155<V_{LSR}<-75$
\kms, except for the case of \ion{O}{6}, which appears to extend to
larger negative velocity.  The feature centered at
$V_{LSR}\approx-215$ \kms\ in the \ion{O}{6} profile is 
Lyman (6-0) [L6-0] P(3) \Htwo\ absorption.  Owing to low signal to noise, 
the $J=3$ \Htwo\ profiles in this dataset are sufficiently weak that we 
are unable to model an
expected profile of the contamination in the \ion{O}{6} profile.
Based on absorption due to the stronger L4-0 R(3) line, we can
establish with certainty that the absorption in the \ion{O}{6} profile
is uncontaminated by \Htwo\ redward of $V_{LSR}=-180$ \kms, and that
absorption can be attributed to \ion{O}{6}.  Thus, Complex C
\ion{O}{6} in this sight line extends beyond low-ion absorption, and
we integrate over $-180<V_{LSR}<-75$ \kms\ for its measurement.
        
Complex C equivalent width measurements for this sight line are shown
in Table 8.  Because of the limited number of line measurements, we
are unable to empirically derive a curve-of-growth.  
Previous analysis suggested that saturation can be an issue with the
\ion{O}{1} $\lambda1039.230$ and \ion{Fe}{2} $\lambda1144.938$ lines,
and some saturation may be indicated by their profiles in Figure 7. 
Therefeore, the apparent optical depth method may not accurately determine
\ion{O}{1} and \ion{Fe}{2} column density.  We determine their range
of possible column densities by considering the range of doppler
parameters $10.2\leq b\leq 18.0$ \kms\ for curve-of-growth fits to
\ion{O}{1} and \ion{Fe}{2} lines in other Complex C sight lines.
These column density ranges are shown in brackets in Table 8.  
All other column densities are calculated by the apparent optical
depth method, including those from \ion{Si}{2} $\lambda1020.70$ and 
\ion{O}{6} $\lambda1031.926$, which are clearly optically thin lines.

\section{DISCUSSION}

\subsection{Revised Solar Abundances and Atomic Data}

With the measured column densities for the Complex C sight lines, we 
are now able to measure abundances, relative to the recently revised 
solar (photospheric) abundances (Asplund, Grevesse, \& Sauval 2005,
hereafter AGS05).  In CSG03, we used solar 
abundances from Grevesse \& Sauval (1998). The largest changes 
in our current analysis are: (Ar/H)$_{\odot} = 10^{-5.82}$; 
(N/H)$_{\odot} = 10^{-4.22}$; and (O/H)$_{\odot} = 10^{-3.34}$.  
Substantial changes occurred for argon ($\Delta A_{\rm Ar} = -0.22$ dex),
oxygen ($\Delta A_{\rm O} = -0.17$ dex), and nitrogen
($\Delta A_{\rm Ar} = -0.14$ dex).  The abundances changed less
significantly for silicon ($\Delta A_{\rm Si} = -0.04$ dex),
sulfur ($\Delta A_{\rm S} = -0.06$ dex), and iron
($\Delta A_{\rm Fe} = -0.05$ dex).

Most of these abundances are only accurate to $\pm 0.05$ (dex).
However, the downward revisions in solar CNO abundances and of Ar, which
is tied indirectly to oxygen, remain controversial.  The revisions arise
primarily from the application of time-dependent, non-LTE, 3D hydrodynamical
models of the solar atmosphere, which have decreased the metal abundance
in the solar convection zone by almost a factor of two, compared to previous
compilations (Anders \& Grevesse 1989; Grevesse \& Sauval 1998).
For example, the new solar abundances by mass of hydrogen, helium, and
metals have $X = 0.7392$, $Y = 0.2486$, and $Z = 0.0122$ (AGS05), and
thus have much lower $Z/X = 0.0165$ compared to the previously recommended
value of $Z/X = 0.0275$ (Anders \& Grevesse 1989).  These lower
metallicities are in poor agreement with helioseismological solar profiles
of sound speed and density (Bahcall \etal\ 2005).  They also disagree with
C and O abundances derived from solar CO modeling.  From their CO models,
Ayres, Plymate, \& Keller (2006) recommend a higher oxygen abundance,
$(7 \pm 1) \times 10^{-4}$, a factor of 0.185 dex larger than that
recommended by AGS05.   We mention these issues, because it is
possible that the reference solar abundance may change yet again.
Therefore, one must regard the [O/H] metallicities in Complex C as
relative values, which could drop by 0.10--0.15 dex if the standard solar
oxygen abundance increases to accommodate the constraints from
helioseismological data.

\thispagestyle{empty}
\begin{deluxetable*}{lccccccccc}
\tablecolumns{11}
\tablewidth{0pc}
\tablecaption{COMPLEX C ABUNDANCE MEASUREMENTS\tablenotemark{a}} 
\tablehead{\colhead{Sight Line} & \colhead{log \NHI\ (cm$^{-2}$)} & \colhead{\ion{O}{1}} & \colhead{\ion{N}{1}} & \colhead{\ion{Si}{2}} & \colhead{\ion{Fe}{2}} & \colhead{\ion{S}{2}} & \colhead{\ion{Ar}{1}} }
\startdata
3C 351\tablenotemark{b}      & $18.62^{+0.03}_{-0.03}$ & \scriptsize $[-0.91^{+0.06}_{-0.07},$ & $<$$-1.08$ & $-0.35^{+0.12}_{-0.12}$ & $-0.29^{+0.08}_{-0.10}$ & $<0.12$ & ...\\
                             &                         & \scriptsize $-0.67^{+0.08}_{-0.08}]$  &            &                         &                         &         & ... \\
HS 1543+5921\tablenotemark{b} & $19.67^{+0.05}_{-0.07}$ & ...                     & \scriptsize $[-1.22^{+0.17}_{-0.17},$ & ...  & ...     & $-0.51^{+0.10}_{-0.12}$ & ... \\
                              &                         &                         & \scriptsize $-0.53^{+0.50}_{-0.43}]$ &      &         &                         &     \\
Mrk 205                      & $18.11^{+0.09}_{-0.11}$ & $-0.78^{+0.12}_{-0.13}$ & $<$$-0.63$              & $-0.51^{+0.16}_{-0.15}$  & $<$$-0.13$              & $<0.57$  & $<0.60$\\
Mrk 279                      & $19.27^{+0.08}_{-0.11}$ & $-0.95^{+0.14}_{-0.13}$ & $-1.83^{+0.14}_{-0.17}$ & $-0.64^{+0.11}_{-0.14}$  & $-0.87^{+0.14}_{-0.16}$ & $-0.72^{+0.14}_{-0.19}$ & $<$$-0.76$\\
Mrk 290                      & $20.12^{+0.02}_{-0.02}$ & $-1.03^{+0.29}_{-0.16}$ & $-1.67^{+0.18}_{-0.21}$ & $-0.70^{+0.19}_{-0.14}$ & $-1.16^{+0.24}_{-0.22}$ & $-1.02^{+0.17}_{-0.21}$ & $<$$-1.15$\\ 
Mrk 817                      & $19.51^{+0.01}_{-0.01}$ & $-0.54^{+0.16}_{-0.09}$ & $<$$-1.17$              & $-0.56^{+0.09}_{-0.09}$ & $-0.68^{+0.08}_{-0.09}$ & $-0.31^{+0.09}_{-0.08}$ & $<$$-1.03$\\
Mrk 876- C. 1                & $19.30^{+0.03}_{-0.04}$ & $-0.70^{+0.18}_{-0.15}$ & $-1.18^{+0.14}_{-0.12}$ & $-0.25^{+0.12}_{-0.11}$  & $-0.39^{+0.09}_{-0.09}$ & $<$$-0.10$              & $<$$-0.69$\\
Mrk 876- C. 2                & $18.72^{+0.12}_{-0.17}$ & $-0.59^{+0.22}_{-0.21}$ & $<$$-1.13$              & $-0.38^{+0.16}_{-0.20}$ & $-0.45^{+0.20}_{-0.22}$ & $<0.40$                 & $<$$-0.07$\\ 
PG 1259+593                  & $19.95^{+0.01}_{-0.01}$ & $-1.00^{+0.21}_{-0.18}$ & $-1.81^{+0.25}_{-0.18}$ & $-0.80^{+0.22}_{-0.18}$ & $-1.01^{+0.12}_{-0.11}$ & $-0.71^{+0.10}_{-0.10}$ & $-1.05^{+0.12}_{-0.14}$\\
PG 1351+640                  & $19.83^{+0.02}_{-0.02}$ & $<$$-0.42$              & $<$$-0.75$              & $-0.44^{+0.31}_{-0.20}$  & $-0.50^{+0.26}_{-0.18}$ & $<$$-0.47$              & $<$$-0.60$\\
PG 1626+554\tablenotemark{b} & $19.43^{+0.02}_{-0.02}$ & \scriptsize $[-0.79^{+0.10}_{-0.10},$ & $<$$-0.72$ & $-0.38^{+0.12}_{-0.17}$ & \scriptsize $[-0.69^{+0.11}_{-0.12},$ & ... & $<$$-0.36$\\
                             &                         & \scriptsize $-0.41^{+0.23}_{-0.20}]$ &            &                         & \scriptsize $-0.43^{+0.23}_{-0.20}]$   &     &           \\
\enddata
\tablenotetext{a}{The abundance of element X is calculated relative to \ion{H}{1}.  
The 0.1 dex systematic error due to the beam-size mismatch between \ion{H}{1} 21-cm 
data and the quasar absorption spectra is not listed.} 
\tablenotetext{b}{Column density range for quantities in brackets is calculated 
assuming doppler parameter $10.2<b<18.0$ \kms.}
\end{deluxetable*}

Additional small changes in metal-ion column densities arise from new
curves of growth based on updated oscillator strengths for selected 
lines (Morton 2003).
Significant changes between CSG03 and the current paper occurred for
O~I, where $f(\lambda 1302)$ decreased from 0.052 to 0.048, and
Fe~II where $f(\lambda 1144$) decreased from 0.106 to 0.087,
$f(\lambda 1143$) increased from 0.0177 to 0.0192, and
$f(\lambda 1122$) increased from 0.020 to 0.029.

We have gone back to those three sight lines and re-fitted curves of 
growth using the new $f$-values, recalculated column densities,
and calculated abundances relative to the new solar values of AGS05.
The wavelengths covered by \HST-STIS E140M data toward 3C~351 include only
one strong \ion{O}{1} line, which does not allow a curve
of growth fit.  Given the strength of the \ion{O}{1} line, unresolved
saturation may be an issue.  We therefore take the same approach as for
PG 1626+554, establishing possible \ion{O}{1} column
densities for a range of $b$-values from $b=10.2$ to $b=18.0$ \kms.
Other AOD column densities toward 3C~351 reported in Tripp \etal\ (2003)
are scaled to the new Morton (2003) $f$-values, and line upper limits are
converted from 4$\sigma$ to 3$\sigma$.  Abundance measurements for all
Complex C sight lines are shown in Table 9.  Note that the abundances
in Table 9 do not include the 0.1 dex systematic error associated with
the beam-size mismatch.

\subsection{Abundances in Complex C}

In order to properly analyze the abundances in Complex C, it is
important to include measurements from all available Complex C sight
lines.  There are four additional Complex C sight lines, not presented
here, which have been thoroughly analyzed in previous studies.  CSG03
analyzed the additional Complex C sight lines Mrk~817, PG~1259+593, and 
PG~1351+640.  Tripp \etal\ (2003) included equivalent
width and apparent-optical-depth column density measurements toward
the 3C 351 sight line.  

Using this dataset, we find that Complex C has metallicity ranging from 
0.09 to 0.29 $Z_{\sun}$, based on the [\ion{O}{1}/\ion{H}{1}] abundance. 
This range is based strictly on the seven Complex C components in six 
sight lines, where we can empirically fit a curve of growth.  Two other 
sightlines have less reliable values of [\ion{O}{1}/\ion{H}{1}], owing 
to a range of possible 
doppler parameters for the curve of growth.  The allowed range 
of metallicities towards 3C 351 is consistent with the more reliable 
data, while the range of \ion{O}{1} abundances towards PG~1626+554 
allows a metallicity as high as 0.39 $Z_{\sun}$ if $b=10.2$ \kms.  
Our reliably measured range, $Z/Z_{\sun}=0.09-0.29$, indicates a 
factor-of-three variation in metal enrichment across Complex C.  
The mean \ion{O}{1} abundance of those 7 components yields an
average Complex C metallicity of $0.17~Z_{\sun}$, with a 
column-density-weighted mean of $0.13~Z_{\sun}$.  As with the conclusions 
of previous Complex C studies (CSG03; Tripp \etal\ 2003), these
metallicities indicate that Complex C consists of more than just
primordial material, with enriched material as well, most likely of
Galactic origin (or expelled from Local group dwarfs).  
As for the range in metallicity, 
certain regions of Complex C appear to be more highly mixed with
enriched gas than others.  
The $1\sigma$ errors on these measurements are typically within a
factor of 1.5, and the beam-size systematic error is within a factor
of 1.25.  It is unlikely that that these results are consistent with
a uniform metallicity for Complex C. 

In order to trace possible signatures of gas mixing, we looked for 
trends in [\ion{O}{1}/\ion{H}{1}].  One possible signature
of gas mixing could be a dependence of
[\ion{O}{1}/\ion{H}{1}] on \ion{H}{1} column density.  Sight lines
passing through low-\NHI\ regions could trace areas where gas mixing is
more efficient, whereas high-\NHI\ regions mark cloud cores, where
mixing may not be as effective.  Such a scenario is consistent with the
3 [\ion{O}{1}/\ion{H}{1}] measurements presented in CSG03.  Figure 8
shows a plot of [\ion{O}{1}/\ion{H}{1}] vs.\ log~\NHI\ for each of the
10 Complex C components presented here.  The Mrk 290 measurement 
encompasses two velocity components.  One of the components has much 
larger \NHI\ and dominates the abundance measurement; we therefore
consider it as effectively one component.  Figure 8 does not include 
the 0.1 dex systematic error from the beam-size mismatch.  

Because of its importance as an indicator of mixing, we now examine 
Figure 8 for potential metallicity dependence on \NHI.  The two highest 
column density components, toward PG~1259+593 and Mrk~290, do have the 
lowest metallicity of $\sim10$\% solar.  However, owing to the error bars,
the statistical significance of a trend is weak.  If we use the 
column-density weighted mean metallicity (0.13 $Z_{\sun}$), we find  
a 11\% probability that the observed variance in [\ion{O}{1}/\ion{H}{1}] vs.\ 
log~\NHI\ occurs purely by chance.  Therefore, any slope to the relation
of metallicity with \NHI\ is present only at the $2 \sigma$ level.    
In order to properly assess the metallicity dependence on \NHI, it would 
be desirable to gain more precise measurements along more sight lines,
using additional data from \FUSE\ or from the {\it Cosmic Origins 
Spectrograph} scheduled for 2008 installation on \HST, in order to 
reduce the error bars on O~I metallicity.   

Another possible tracer of gas mixing would be a metallicity dependence 
on Galactic latitude.  Assuming that higher latitude regions of
Complex C are at larger distance from the Galactic plane, it is
possible that low-latitude sight lines may pierce regions that are
closer to the disk and more highly enriched by solar-metallicity
fountain material.  Referring to a sight-line map of Complex C
(e.g., Figure 1 of CSG03), such a scenario is not borne out by the
data.  Although the two lowest metallicity sight lines are at
relatively high latitude, the highest-metallicity sight line, toward
Mrk 817, is also at high latitude.  Although, the magnitude and
variation in sight-line metallicity indicates different degrees of
primordial and enriched gas mixing throughout Complex C, the signature
of mixing versus Galactic latitude is not apparent.
 
\begin{figure*}
\figurenum{8}
\epsscale{1}
\plotone{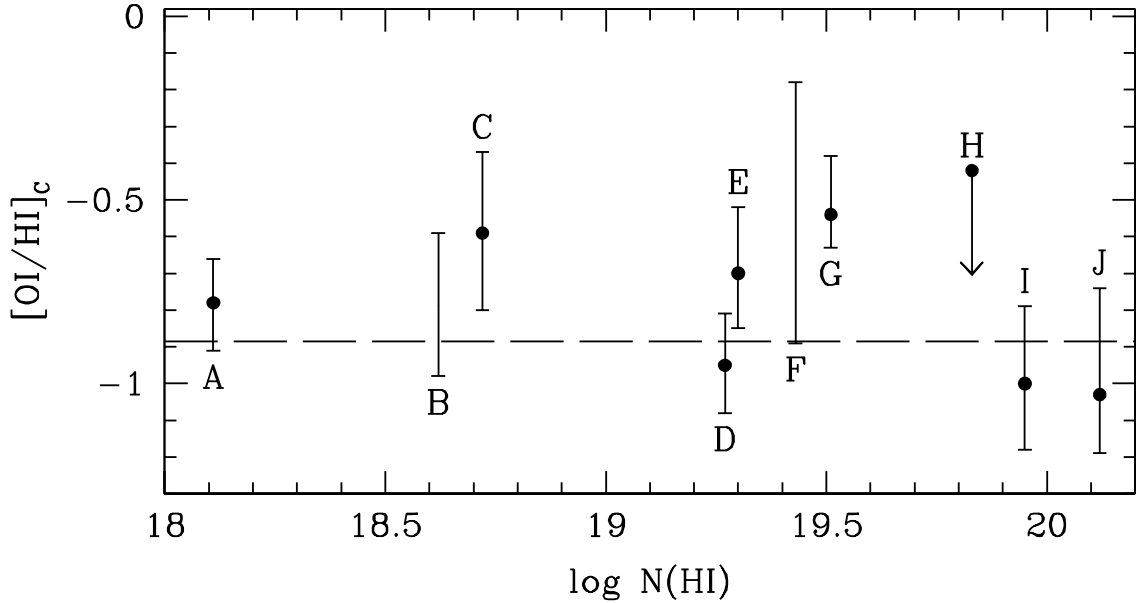}
\caption{Plots of [\ion{O}{1}/\ion{H}{1}] vs.\ \ion{H}{1}\ column
density for the 9 Complex-C sight lines (10 velocity components)
discussed in this paper.  Seven components have reliably measured
[O~I/H~I].  Dotted line shows the column density weighted mean of 
$0.13 Z_{\odot}$.  [\ion{O}{1}/\ion{H}{1}] measurements towards 3C~351
and PG~1626+554 are illustrated by a range that includes the $1\sigma$
error bars on the minimum and maximum values arising from the
uncertain range in $b$.  Letters next to data points identify the
Complex C components as follows: A) Mrk 205; B) 3C 351; C) Mrk 876,
Component 2; D) Mrk 279; E) Mrk 876, Component 1; F) PG 1626+554; G)
Mrk 817; H) PG 1351+640; I) PG 1259+593; J) Mrk 290.}
\end{figure*}

\subsection{Relative Abundances and Nucleosynthetic History} 

An additional issue worth consideration is the nucleosynthetic
history of the cloud as revealed by relative abundance patterns.
Previous Complex C studies have found evidence of $\alpha$-element (S,
Si, O) enhancement and nitrogen depletion, indicating that Complex C
has been enriched primarily by metals produced by Type II SNe (Richter
\etal\ 2001; CSG03; Tripp \etal\ 2003).  These topics can be
investigated more fully, now that column densities of \ion{O}{1} and
\ion{N}{1} have been measured for several more sight lines.  The
consideration of relative abundances instead of absolute abundances
has the additional advantage that the problem of UV/radio beam-size 
mismatch is not applicable.  In order to consider relative abundances, 
we have calculated the column-density-weighted abundances of \ion{S}{2},
\ion{Si}{2}, \ion{Fe}{2}, and \ion{N}{1}, relative to \ion{O}{1}.  
We only include a sight line's relative abundance if abundances of 
both of the considered ion species have been measured.  These totals are:   
[\ion{S}{2}/\ion{O}{1}] = $0.16^{+0.16}_{-0.13}$,
[\ion{Si}{2}/\ion{O}{1}] = $0.27^{+0.19}_{-0.13}$,
[\ion{Fe}{2}/\ion{O}{1}] = $-0.03^{+0.16}_{-0.12}$,
and [\ion{N}{1}/\ion{O}{1}] = $-0.69^{+0.21}_{-0.17}$.
The [\ion{N}{1}/\ion{O}{1}] ratio is based on only the four sight lines in 
which \ion{N}{1} lines were successfully measured.  The cases with only upper
limits are also consistent with a significantly depleted 
\ion{N}{1} relative abundance.   Toward Mrk 817 and in Component 2 toward 
Mrk~876, we find Complex C abundances of [\ion{N}{1}/\ion{O}{1}] $< -0.63$ 
and [\ion{N}{1}/\ion{O}{1}] $< -0.54$, respectively.

In order to determine relative elemental abundances from ion
abundances, one must make ionization corrections, defined for example as   
[S/H] = [S~II/H~I] + (correction), where logarithmic quantities
are implied.  Such corrections are uncertain, because they 
depend on the gas density and ionizing radiation field. 
CSG03 calculated such corrections, using the photoionization 
code Cloudy (Ferland \etal\ 1998; see Figure 16 from CSG03).
The corrections were negligible for \ion{O}{1} and \ion{N}{1}, which 
are coupled to \ion{H}{1} through charge exchange.  Using version 06.02 
of Cloudy, we have updated these calculations, constructing
a new grid of models with input parameters similar to those
described in CSG03, but with relative abundances consistent with AGS05. 
The results for O~I and N~I, as well as Si~II, S~II, and Fe~II are
similar to that shown in CSG03 for log~\NHI\ down to $19$.
The new sightline, toward MRK 205, for which log~\NHI\ = 18.1,
will have larger ionization corrections but the limited data
available here will not impact our column density weighted results.
The doubly ionized 
species, \ion{Si}{2}, \ion{S}{2}, and \ion{Fe}{2}, occur in both
\ion{H}{1} and \ion{H}{2} regions, and ionization corrections are
necessary to properly compare relative abundances involving these 
singly-ionized species.  For 
Si, S, and Fe, the sign of the correction is negative
(abundances of Si, S, and Fe are {\it less} than inferred
from Si~II/H~I, S~II/H~I, Fe~II/H~I) and the 
magnitude decreases at larger \NHI.  Because the mean relative
abundances are weighted by column density, the ionization
correction is fairly small for \ion{S}{2}, \ion{Si}{2}, and
\ion{Fe}{2}, ranging from 0.1--0.2 dex, with \ion{Fe}{2} requiring
roughly half the logarithmic correction as \ion{Si}{2} and \ion{S}{2}.

Considering a rough ionization correction to the above relative ion
abundances, the $\alpha$-elements (S, Si, and O) have an essentially
solar pattern and are slightly enhanced relative to Fe.  Nitrogen is 
depleted by a factor of 5 from a solar relative abundance pattern.
The enrichment of $\alpha$-elements relative to N and Fe suggests that
the metals were produced primarily by massive stars and injected into
the ISM by Type II SNe.  Nitrogen and iron are typically produced in
longer-lived stars, so an $\alpha$-enhancement is a signature of gas
that has been stripped quickly from the star-forming region.
Material ejected from the disk by vigorous SN activity in a Galactic
fountain, or falling into the disk from galactic winds in Local Group
dwarf galaxies, is consistent with such a scenario.  A plausible
explanation, then, for the observed metallicity and abundance pattern
is that Complex C is mixing with wind/fountain material as it falls 
onto the Milky Way disk.

We now have a fairly complete set of \ion{Ar}{1} upper limits and one
reliable \ion{Ar}{1} detection.  Table 9 shows the \ion{Ar}{1} abundance
results for each Complex C component.  We have only one reliable
Complex C detection, toward PG~1259+593 (CSG03; Sembach \etal\ 2004), where 
[Ar~I/H~I] = $-1.05^{+0.15}_{-0.14}$ (9\% solar) with log~\NHI\ = 19.95
relative to an assumed solar abundance (Ar/H)$_{\odot} = 1.5 \times 10^{-6}$.  
Many of the other Complex C sight lines have lower \NHI, insufficient
to detect the weak Ar~I lines at 1048 and 1066 \AA\ without extremely
long exposures (PG~1259+593 had total exposure $\sim600$ ks).   
However, we should have been able to detect Ar~I toward Mrk~290,
with log~\NHI\ = 20.12, but we only measured an upper limit, 
log~N(Ar~I) $< 13.15$, corresponding to [Ar~I/H~I] $< -1.15$ (7\% solar).     
 
Such low values for [Ar~I/H~I] are not entirely due to metallicity,
as argon probably has a substantial (positive) ionization correction. 
This effect is similar to that seen (Jenkins \etal\ 2000) in the 
local ISM toward four white dwarfs, where they found that Ar~I is deficient 
by 0.4 dex, probably because Ar~I is more easily ionized than H~I.   
This effect probably arises (Sofia \& Jenkins 1998) because Ar~I has 
a large photoionization cross section near threshold (15.755 eV) that 
enhances the abundance in the Ar~II ionization stage.  From our grid of 
photoionization models, we find that a positive ionization correction is 
likely appropriate.  For log~\NHI\ $= 20$, this correction may range from
$0.1$ dex to as much as $0.4$ dex if the volume density is as low as 
$0.01$ cm$^{-3}$.

\subsection{High Ions in Complex C}

In addition to allowing a thorough investigation of the ion abundance
pattern in Complex C, these new data provide the spectral coverage to
analyze highly-ionized species.  The most useful way to study the high
ions is through the comparison of high-ion ratios involving
\ion{C}{4}, \ion{Si}{4}, \ion{N}{5}, and \ion{O}{6}.  In order to
carry out such an analysis, we require that Complex C sight lines
have both STIS E140M and \FUSE\ data.  Until the acquisition of the new
data presented in this paper, the only sight line with both \FUSE\ and
STIS E140M data was PG 1259+593.  F04 thoroughly
investigated the high-ions in Complex C toward PG 1259+593 and found
that the values are consistent with the high ions arising through
collisional ionization at the interface of Complex C and a surrounding
hotter medium.  With the new STIS E140M data for Mrk 279 and Mrk 876, 
we can now investigate Complex C high ions in three sight lines.  

Table 10 shows the logarithmic column density ratios involving
\ion{C}{4}, \ion{Si}{4}, \ion{N}{5}, and \ion{O}{6} for these three 
sight lines.  Included is the PG 1259+593 measurement from CSG04.  The
high ion absorption lines are typically weak and show little evidence
of saturation.  The $W_{\lambda}$ measurements of the \ion{C}{4} and
\ion{Si}{4} doublets toward Mrk 876 confirm an absence of saturation.
In the cases of Mrk 279 and PG 1259+593, the weaker lines of the
\ion{C}{4} and \ion{Si}{4} doublets cannot be measured.  Owing to their
lack of saturation and better detectability, we use the stronger lines
of these doublets to calculate column density ratios.
The high-ion ratios are remarkably similar for each of the three sight
lines.  Although \ion{N}{5} cannot be measured, the values for
[\ion{C}{4}/\ion{O}{6}] and [\ion{Si}{4}/\ion{O}{6}] vary by no more
than 0.1 dex.  

These measured ratios can be compared to ionization models.
Photoionization models used to fit high ions in HVCs require a large
ionization parameter, and thus low gas density and size $\sim100$ kpc
(F04, CSG04).  With an approximate upper limit on the distance to Complex C
at $\lesssim25$ kpc (Wakker \etal\ 1999) and, in turn, a size no larger
than a few kpc, photoionization can be ruled out for producing the
high ions in Complex~C.   Collisional ionization can produce copious
amounts of high ions, while photoionization with a more reasonable
ionization parameter can produce the low ions.  F04 consider several
common collisional ionization models for the case of a low-metallicity
cloud.  They present corrected high-ion ratio predictions for these
models, using C, O, Si, and N abundances appropriate for Complex C.
The abundances they use are based on the PG 1259+593 results, which
differ from what we measure towards Mrk 279 and Mrk 876, and from 
the PG 1259+593 results presented here, although they agree to
within 0.2 dex.  This uncertainty in the input to the corrected models
is small compared to the predicted ranges of high-ion ratios, and thus
does not significantly affect the analysis.

\begin{deluxetable}{lccc}
\tablecolumns{4}
\tablewidth{0pc}
\tablecaption{COMPLEX C LOGARITHMIC COLUMN DENSITY RATIOS\tablenotemark{a}
\label{t11}}
\tablehead{
\colhead{Sight Line} & \colhead{[\ion{C}{4}/\ion{O}{6}]} & \colhead{[\ion{Si}{4}/\ion{O}{6}]} & \colhead{[\ion{N}{5}/\ion{O}{6}]}}
\startdata
Mrk 279     & $-0.36^{+0.06}_{-0.06}$ & $-0.93^{+0.06}_{-0.06}$ & $<$$-0.82$ \\
Mrk 876     & $-0.40^{+0.04}_{-0.04}$ & $-0.92^{+0.04}_{-0.03}$ & $<$$-0.88$ \\
PG 1259+593 & $-0.31^{+0.08}_{-0.09}$ & $-0.90^{+0.11}_{-0.14}$ & $<$$-0.38$ \\
\enddata
\tablenotetext{a}{We present high-ion ratios for only these three sightlines 
for which \FUSE\ and \HST\ data provide wavelength coverage of the UV lines for
O~VI (1032~\AA), C~IV (1548~\AA), N~V (1238~\AA), and Si~IV (1394~\AA).}  
\end{deluxetable}

Comparing the high-ion measurements to the corrected models of F04, we
can assess possible ionization sources.  We find that the observed
high ion ratios can be explained by subsolar abundance models of shock
ionization (Dopita \& Sutherland 1996) or turbulent mixing layers
(Slavin, Shull, \& Begelman 1993).  The conductive interface model
(Borkowski, Balbus, \& Fristrom 1990) produces lower ratios of
[\ion{Si}{4}/\ion{O}{6}] than observed.  However \ion{Si}{3} is more
easily photoionized (ionization potential = 33.5 eV) than the other
high ions.  If the bulk of the \ion{Si}{4} is produced through
photoionization, then ionization on conductive interfaces could
explain the production of the other high ions.  Although one can
relax the size requirements on collisionally ionized high ions, 
one still cannot violate the mass constraints (CSG05) of placing
the system of HVC O~VI at large distances (100~kpc to Mpc scales).  

Although one cannot prove that high-velocity O~VI is co-spatial 
with the low ions, its kinematic association suggests a connection.  
In our previous papers (CSG04, CSG05), we discussed hybrid ionization 
models, in which high- and low-ions are produced by a combination of 
photoionization and collisional ionization, as in bow shocks and other 
interfaces.  Collisional ionization by extended ($D > 100$~kpc)
low-density gas requires too much mass and is inconsistent with
halo scale lengths.  Each of the viable collisional ionization sources 
requires interaction between Complex C and a surrounding environment.  
For turbulent mixing and conductive heating, the interaction is
from immersion in a surrounding hotter medium, possibly an extended
Galactic corona.  Shock ionization may occur in a bowshock as the
Complex plunges through the halo towards the disk.  The shock
possibility was explored (CSG04, CSG05) as a method of producing 
the high ions in highly ionized HVCs.  The line profile kinematics
support the scenario of the high ions tracing the outer envelope or
leading edge of Complex C.  High-ions often extend over a larger
velocity range than the low ions, suggesting that the low ions trace
denser, photionized gas at the cloud core.  Such an effect is seen in
the data for Mrk 279, Mrk 290, and PG 1626+554. 

The remarkable kinematics of the high ions toward Mrk 876 suggest a 
decelerated bowshock, since the high-velocity \ion{C}{4} and \ion{Si}{4} 
profiles peak at $+30$ \kms\ from the
low ions.  Further, the high ions are well blended with lower velocity
gas, whereas the low ions are not, suggesting a possible mixing of hot
gas in Complex C and the Galactic halo.  The high ions toward Mrk 876
may trace a decelerating shocked medium at the leading edge of the
cloud as it falls through and interacts with the Galactic halo.

Analysis of the high ions in Complex C can also be used to understand
the nature of the highly ionized HVCs by comparing to the high ion
ratios (\ion{C}{4}, \ion{Si}{4}, \ion{N}{5}, \ion{O}{6}) 
presented by CSG05.  Although they show a larger spread, the
high-ion ratios for the highly ionized HVCs that are
detected in singly-ionized species
are similar to the values presented in Table 11, suggesting a similar
ionization source.  Only the highly ionized HVCs 
detected in \ion{O}{6} alone show a marked deviation from what is
observed in Complex C sight lines.  The characterization of the
Complex C high ions in this work adds further to the proposal 
by CSG05 that the majority of highly ionized HVCs represent low-\NHI\
analogs to the large HVCs like Complex C.

\section{CONCLUSIONS}

In this paper, we assembled the latest dataset of \FUSE\ and
\HST-STIS spectra of 10 Complex C sight lines.  Including new data 
for several sightlines, and reanalyzing previous data, we derived 
abundance ratios for 11 Complex C velocity components detected in 
UV absorption lines.  We also revised or updated the H~I (21-cm) 
column densities, solar abundances, and absorption oscillator strengths. 
Values of log~\NHI\ from 21-cm emission typically changed by 
$\leq 0.03$ between our 2003 analysis, based on Wakker \etal\ (2001), 
and our current analysis, based on Wakker \etal\ (2003).  

Using the new data, with revised values of \NHI, solar abundances, and 
absorption-line $f$-values, we arrive at the following conclusions:

\begin{enumerate}

\item {\it Metallicity.} We measure the ratio of
[\ion{O}{1}/\ion{H}{1}] with curve-of-growth techniques in seven Complex~C 
components and constrain the ratio in two other components.  The
\ion{O}{1} abundance has been argued to be an accurate tracer of gas
metallicity based on charge-exchange coupling with \ion{H}{1}.
Our results significantly improve the statistics of Complex C
metallicity measurements from previous studies.  Based on 
[\ion{O}{1}/\ion{H}{1}], we find that the metallicity of
Complex C varies from 0.09 to 0.29 $Z_{\sun}$.  This factor-of-three
variance in metallicity is unlikely to be attributed to systematic
uncertainties such as mismatched radio-emission and UV-absorption beam
sizes.  The column-density weighted mean metallicity of the sample is
$Z = 0.13~Z_{\sun}$, with a possible but unconfirmed ($2 \sigma$) 
correlation of lower 
metallicity along sightlines with higher \NHI.  However, with our
current data, the variations in metallicity and their error bars make 
it difficult to prove this trend.  Better data might demonstrate more 
efficient mixing of infalling gas with lower-\NHI\ gas in the Complex-C 
envelopes, compared to lower-metallciity cloud cores. 

\item {\it Relative Abundances.}  The abundance ratio
[\ion{N}{1}/\ion{H}{1}] indicates a significant $\alpha$-element
enrichment in Complex C gas.  We measure [\ion{N}{1}/\ion{H}{1}] over
the range 0.01 to 0.07 (N/H)$_{\sun}$ in four sight lines.  
Comparing to the \ion{O}{1} abundance, we measure a column density 
weighted mean relative abundance of 0.20 (N/O)$_{\sun}$.
Further, Fe is slightly depleted relative to the $\alpha$-elements (O,
S, Si).  The enrichment of $\alpha$-elements relative to N and Fe
suggests metal production by massive stars, with subsequent
contamination of the ISM by Type II SNe (gas in the Galactic fountain
or expelled from Local Group dwarfs).  Such a nucleosynthetic
history is consistent with Complex C metal enrichment by fountain or
wind material, as the object interacts with the Galactic halo.

\item {\it Highly Ionized Gas.} The three sight lines with both 
\FUSE\ and \HST-STIS data allow spectral coverage of the key high
ions \ion{C}{4}, \ion{Si}{4}, \ion{N}{5}, and \ion{O}{6}.  The high ion 
column density ratios are nearly identical for each of three
sight lines, and are consistent with the presence of hot
$10^{5}-10^{6}$ K gas at the cloud interface.  The highly ionized
ratios are adequately explained by either ionization by shocks or an
interaction with a surrounding medium via turbulent mixing or
conductive interfaces.  The high ion ratios of the \ion{H}{1}-detected
Complex C are similar to those of the highly ionized HVCs, which are
detected only in UV absorption spectra.  The possible similarity in
ionization mechanism suggests each of these classes of HVCs may trace
the same phenomenon, though at different ends of the HVC column
density distribution.

\end{enumerate}

\acknowledgments

We acknowledge support from NASA/FUSE grant NNG04GO36G and STScI 
archival grant AR-10645.02-A at the University of Colorado.  The 
FUSE data were obtained by the
Guaranteed Time Team for the NASA-CNES-CSA mission operated by the
Johns Hopkins University.  Financial support to U.S. participants was
provided by NASA contract NAS5-32985.


\begin{references}

\reference{}Anders, E., \& Grevesse, N. 1989, Geochim.\ Cosmochim.\ Acta,
   53, 197

\reference{}Asplund, M., Grevesse, N., \& Sauval, A. J. 2005, in ASP Conf.\ 
   Ser.\ 336, Cosmic Abundances as Records of Stellar Evolution and 
   Nucleosynthesis, eds. T. G. Barnes III and F. N. Bash (San Francisco: ASP), 
   25 (AGS05)

\reference{}Ayres, T. R., Plymate, C., \& Keller, C. U. 2006, \apjs, 165, 618

\reference{}Bahcall, J. N., Basu, S., Pinsonneault, M., \& Serenelli, A. M.  
    2005, \apj, 618, 1049

\reference{}Bland-Hawthorn, J., \& Putman, M. E. 2001, in ASP Conf.\ 
   Ser.\ 240, Gas and Galaxy Evolution, eds. J. E. Hibbard, M. Rupen, 
   and J. H. van Gorkom (San Francisco: ASP), 369

\reference{}Borkowski, K. J., Balbus, S. A., \& Fristrom, C. C. 1990, 
   \apj, 355, 501

\reference{}Bowen, D. V., Jenkins, E. B., Pettini, M., \& Tripp, T. M. 2005, 
   \apj, 635, 880

\reference{}Bregman, J. N. 1980, \apj, 236, 577

\reference{}Collins, J. A., Shull, J. M., \& Giroux, M. L. 2003, \apj, 
   585, 336 (CSG03)

\reference{}Collins, J. A., Shull, J. M., \& Giroux, M. L. 2004, \apj, 
   605, 216 (CSG04)

\reference{}Collins, J. A., Shull, J. M., \& Giroux, M. L. 2005, \apj, 
   623, 196 (CSG05)

\reference{}Dopita, M. A., \& Sutherland, R. S. 1996, \apjs, 102, 161

\reference{}Ferland, G. J., Korista, K. T., Verner, D. A., Ferguson, J. W.,
   Kingdon, J. B., \& Verner, E. M. 1998, PASP, 110, 761

\reference{}Fox, A. J., Savage, B. D., Wakker, B. P., Richter, P., 
   Sembach, K. R., \& Tripp, T. M. 2004, \apj, 602, 737 (F04)

\reference{}Gibson, B. K., Giroux, M. L., Penton, S. V., Stocke, J. T., 
   Shull, J. M., \& Tumlinson, J. 2001, \aj, 122, 3280

\reference{}Gillmon, K., Shull, J. M., Tumlinson, J., \& Danforth, C. 2006, 
     \apj, 636, 891 

\reference{}Grevesse, N., \& Sauval, A. J. 1998, SSRv, 85, 161

\reference{}Hartmann, D., \& Burton, W. B. 1997, Atlas of Galactic Neutral 
    Hydrogen (Cambridge: Cambridge Univ. Press)

\reference{}Henry, R. B. C., Edmunds, M. G., K\"{o}ppen, J. 2000, \apj, 
   541, 660

\reference{}Jenkins, E. B., \etal\ 2000, \apj, 538, L81  

\reference{}Moos, H. W., \etal\ 2000, \apj, 538, L1

\reference{}Morton, D. C.,  2003, \apjs, 149, 205

\reference{}Murphy, E. M., \etal\ 2000, \apj, 538, L35

\reference{}Nicastro, F., \etal\ 2002, \apj, 573, 157

\reference{}Nicastro, F., \etal\ 2003, \nat, 421, 719

\reference{}Pagel, B. E. J. 1994, in The Formation and Evolution of 
   Galaxies, ed. C. Munez-Tun\'{o}n \& F S\'{a}nchez (Cambridge: 
   Cambridge Univ. Press), 149

\reference{}Pettini, M. 2004, in Cosmochemistry: The Melting Pot of 
   Elements, eds. C. Esteban \etal\ (Cambridge: Cambridge Univ. Press), 257

\reference{}Richter, P., et al. 2001, \apj, 559, 318

\reference{}Sahnow, D. J., et al. 2000, \apj, 538, L7

\reference{}Savage, B. D., \& Sembach, K. R. 1991, \apj, 379, 245

\reference{}Sembach, K. R., Savage, B. D., Lu, L., \& Murphy, E. M. 1999,
   \apj, 515, 108

\reference{}Sembach, K. R., \etal\ 2003, \apjs, 146, 165

\reference{}Sembach, K. R., \etal\ 2004, \apjs, 150, 387

\reference{}Shapiro, P. R., \& Field, G. B. 1976, \apj, 205, 762

\reference{}Slavin, J. D., Shull, J. M., \& Begelman, M. C. 1993, 
   \apj, 407, 83

\reference{}Sofia, U. J.., \& Jenkins, E. B. 1998, \apj, 499, 951 

\reference{}Tripp, T. M., \etal\ 2003, \aj, 125, 3122 

\reference{}Wakker, B. P. 2001, \apjs, 136, 463

\reference{}Wakker, B. P., \etal\ 1999, \nat, 402, 388

\reference{}Wakker, B. P., \etal\ 2003, \apjs, 146, 1

\reference{}Wakker, B. P., Kalberla, P. M. W., van Woerden, H., 
   de Boer, K. S., \& Putman, M. E. 2001, \apjs, 136, 537

\reference{}Wakker, B. P., \& van Woerden, H. 1997, \araa, 35, 217

\end{references}
\end{document}